\documentclass{aa}  

\usepackage{newtxtext,newtxmath}

\usepackage[T1]{fontenc}
\usepackage{ae,aecompl}
\usepackage{comment}
\usepackage{aas_macros}
\usepackage[hidelinks]{hyperref}

\usepackage{graphicx}	
\usepackage[export]{adjustbox}

\usepackage{xcolor}

\def\gsim{\;\raise0.3ex\hbox{$>$\kern-0.75em\raise-1.1ex\hbox{$\sim$}}\;}
\def\lsim{\;\raise0.3ex\hbox{$<$\kern-0.75em\raise-1.1ex\hbox{$\sim$}}\;}
\def\gsim{\;\raise0.3ex\hbox{$>$\kern-0.75em\raise-1.1ex\hbox{$\sim$}}\;}

\def\kms{\rm ~km~s^{-1}}

\def\diff{\rm ~cm^2~s^{-1}}
\def \kms {\rm ~km~s^{-1}}

\def\ergs{\rm ~erg~s^{-1}}

\begin{document}

\title{Southern eROSITA bubble as a forward shock and the low-metallicity CGM}
\subtitle{South-east side story}

\author{
Eugene~Churazov \inst{1,2} 
\and
Ildar~I.~Khabibullin \inst{3,1,2,4} 
\and
Andrei~M.~Bykov \inst{5} 
\and
Nikolai~N.~Chugai \inst{6} 
\and
Rashid~A.~Sunyaev \inst{2,1} 
\and
Victor~P.~Utrobin \inst{6,7} 
\and
Igor~I.~Zinchenko \inst{8}
}

\institute{
Max Planck Institute for Astrophysics, Karl-Schwarzschild-Str. 1, D-85741 Garching, Germany 
\and 
Space Research Institute (IKI), Profsoyuznaya 84/32, Moscow 117997, Russia
\and
Rudolf Peierls Centre for Theoretical Physics, Department of Physics, University of Oxford, Clarendon Laboratory, Parks Rd, Oxford, OX1 3PU, United Kingdom
\and
Universitäts-Sternwarte, Fakultät für Physik, Ludwig-Maximilians-Universität München, Scheinerstr.1, 81679 München, Germany
\and
Ioffe Institute, Politekhnicheskaya st. 26, Saint Petersburg 194021, Russia
\and
Institute of Astronomy, Russian Academy of Sciences, 48 Pyatnitskaya str., Moscow 119017, Russia
\and
NRC `Kurchatov Institute', acad. Kurchatov Square 1, Moscow 123182, Russia
\and
Institute of Applied Physics of the Russian Academy of Sciences, 46 Ul'yanov~str., Nizhny Novgorod 603951, Russia
}

\abstract{Unlike the complicated X-ray and radio structure observed in the North Polar Spur area, the South-Eastern part of the eROSITA bubbles can be reasonably well described as a propagating forward shock, plausibly created by the transient energy release at the Galactic Center. In this model, the physical radius of the bubble is $R_{\rm b}\sim 7-8\,{\rm kpc}$ and the age of the outburst is $t_{\rm age}\sim 5-8\,{\rm Myr}$. The latter quantity is plausibly a lower limit on the true age.
The visible segment of the shock front (located at a distance of $\sim 10-12\,{\rm kpc}$ above the Galactic Disk and at a similar distance from the Sun) is currently expanding with the velocity $\sim 700\,{\rm km\,s^{-1}}$ through the gas with density $n_e\sim 3\times 10^{-4}\,{\rm cm^{-3}}$,  and the abundance of heavy elements in this gas is low $Z\sim 0.1-0.2 \times Z_\odot$ (depending on the adopted reference Solar abundances). Unlike constraints derived from the line-of-sight-integrated quantities, these are effectively in situ measurements of the circumgalactic medium (CGM) properties. 
Given the simplifying assumptions used in deriving the density and abundance, we assign a factor of 2 systematic uncertainty to the final estimates.
An eventual decisive test for the shock properties can be provided by the velocity measurements of the X-ray-emitting gas with soft X-ray bolometers. The extended forward shock propagating through low-metallicity gas
is a favorable site to accelerate very high-energy cosmic rays, which 
might contribute to the recently discovered
proton-rich Galactic cosmic ray component at PeV energies impinging on the Earth's atmosphere.
}

\titlerunning{SE eROSITA bubble}

\keywords{X-rays: galaxies -- X-rays: diffuse background -- Galaxy: general}
    
\maketitle

\section{Introduction}
The North Polar Spur/Loop-I (NPS), Fermi Bubbles (FB), and eROSITA Bubbles (EB)\citep[see][for recent reviews]{2023CRPhy..23S...1L,2024A&ARv..32....1S} are prominent large-scale structures in the general direction of the Galactic Center. Whether all these structures are the result of the same phenomenon remains a matter of debate \cite[see, e.g.,][for examples of unified models of FB and NPS/Loop-I]{2015ApJ...811...37M,2015MNRAS.448..328S,2018ApJ...869L..20M,2019MNRAS.482.4813S}. Complexity arises from their large angular size and a possible superposition of unrelated objects. Nevertheless, a model that associates all of them with the energy release in the Galactic Center region \citep[e.g.,][]{1977A&A....60..327S,2020Natur.588..227P,2022NatAs...6..584Y, 2022MNRAS.514.2581M}, remains an attractive solution. 

These three structures are especially prominent in different bands - FBs in gamma-rays, EBs in X-rays, and NPS both at radio frequencies and in X-rays. Here, we focus on the SRG/eROSITA all-sky data, which provide the most detailed view of the entire X-ray structure of EBs and offer a possibility to do detailed spectral analysis in the energy band from 0.3 to a few keV. We took advantage of the all-sky data to sum the signal from a large portion of the south-eastern\footnote{Throughout the paper, we refer to directions with respect to the Galactic coordinate system.} (SE) part of the bubble to get its spectrum and compare it with the predictions of a propagating shock model. This is the main focus of this study, which complements the spectral analysis of selected patches in the direction of the Fermi or eROSITA bubbles with different X-ray satellites \cite[e.g.,][]{2013ApJ...779...57K,2015ApJ...807...77K,2016ApJ...829....9M,2020ApJ...904...54L,2022MNRAS.512.2034Y,2023NatAs...7..799G}.

Yet another interesting question is the contribution of transient events in the Galactic Center region to the most energetic particles that can be produced within the Milky Way. Powerful outflows in starburst galaxies \citep[e.g.][]{2024ARA&A..62..529T} and large, galactic-scale, shocks \citep[e.g.][]{1987ApJ...312..170J} are considered as potential sources of very high energy cosmic rays, neutrino and gamma-rays \citep[e.g.][]{2022Sci...378..538I,2021MNRAS.503.4032A,2023Galax..11...86O,2023A&ARv..31....4R}. The inferred properties of the EB shock suggest that an analogous mechanism, possibly a less powerful version, might operate in the Milky Way too.

\section{Observational data}

The SRG~ X-ray observatory \citep{2021A&A...656A.132S}  was launched on July 13, 2019,  from the Baikonur cosmodrome. It carries two wide-angle grazing-incidence X-ray telescopes, eROSITA \citep{2021A&A...647A...1P} and the Mikhail Pavlinsky ART-XC telescope \citep{2021A&A...650A..42P}, which operate in the overlapping energy bands of 0.2–8 and 4–30 keV, respectively.

We used the data obtained with the eROSITA telescope during four consecutive all-sky surveys.
The initial reduction and processing of the data were performed at IKI using standard routines of the \texttt{eSASS} software \citep{2018SPIE10699E..5GB,2021A&A...647A...1P} and proprietary software developed in the RU eROSITA consortium, while the imaging and spectral analysis were carried out with the background modeling, vignetting, point spread function, and spectral response function calibrations built upon the standard ones via slight modifications motivated by results of calibration and performance verification observations \citep[e.g.,][]{2021A&A...651A..41C,2023MNRAS.521.5536K}.

\section{Images, radial profile, shell model}
\subsection{Image}
\begin{figure*}
\sidecaption
\includegraphics[angle=0,trim=2cm 3.7cm 2cm 3.7cm,clip,width=12cm]{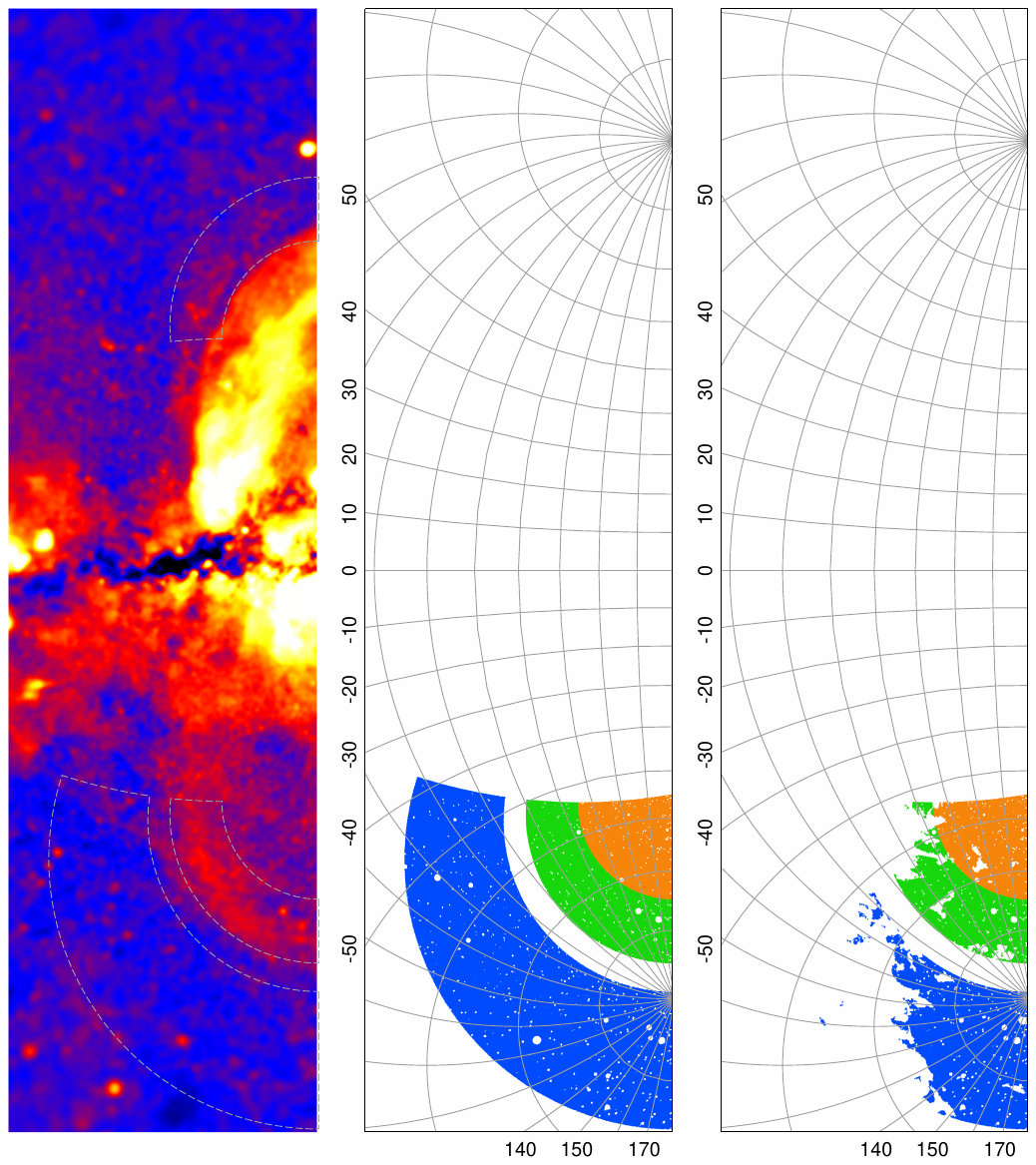}
\caption{X-ray image (0.7-1.05 keV band) of the eROSITA Bubbles in stereographic projection. The Southern bubble has a simpler morphology than the Northern one, resembling a shell characteristic of a propagating quasi-spherical shock. If this is the case, the spectrum could provide constraints on the shock velocity and the distance to the shock.
The two panels on the right show the regions used for the extraction of spectra. The green and blue areas in the middle panel correspond to the "shell" and "background regions, respectively. They are also outlined in the left panel. The right panel shows the same regions after a narrow range of the galactic HI column density ($N_{\rm H}=10^{19}-2.5\times 10^{20}\,{\rm cm^{-2}}$) is considered. This is done in order to ensure that no bias is introduced by  $N_{\rm H}$ variations between the shell and the background regions. Finally, the region outlined by the gray dashed line in the Northern part of the left panel is simply a reflection of the Southern shell to the upper hemisphere to indicate a possible (symmetric) location of the NE counterpart of the SE shell. 
 }
\label{f:sbub_image}
\end{figure*}

Figure~\ref{f:sbub_image} shows the eastern part of the eROSITA bubbles in the 0.7-1.05 keV band in stereographic projection. This projection is useful because both polar regions are visible in one image. The choice of the energy band is motivated by the following consideration. The widespread Milky Way's diffuse emission dominates the sky background at low energies, particularly due to lines of oxygen, nitrogen, and carbon. This emission is highly structured, has large-scale gradients, and suffers from absorption by intervening patches of gas and dust. This emission fades above $\sim 0.7\,{\rm keV}$ and the effects of absorption also decrease. This motivates the choice of this energy as a reasonable lower bound of the energy band used to search for the emission of hotter gas. The upper limit, instead, is set by the desire to get the maximum signal-to-noise ratio for the signal from eROSITA bubbles. The particular choice of $1.05\,{\rm keV}$ as the upper bound ensures that the Ne~X line is included. 

Figure~\ref{f:sbub_image} shows the familiar picture of "eROSITA Bubbles", described in \citep{2020Natur.588..227P}. Here, we focus on the SE part of the Bubbles. Unlike the NE part, which is dominated by the very X-ray-bright and complicated structures co-spatial with the prominent radio-band structure known as the North Polar Spur \citep[see][for review]{2023CRPhy..23S...1L}, the SE region shows a "simple" shell-like structure resembling a bright rim of a supernova remnant. This morphological simplicity justifies a detailed analysis of the SE region in isolation in the context of the propagating shock wave scenario. This scenario is one of the leading explanations for the observed structures associated with AGN or star-formation activity in the Galactic Center region \citep[see][for review]{2024A&ARv..32....1S}.

\subsection{Radial profile across the shell}

It turns out that the outer edge of the SE shell can be reasonably well approximated as a circle with the center at the Galactic coordinates $(l,b)=(0^\circ,-55^\circ)$ and a radius of $32\,{\rm degrees}$. Several concentric shells (wedges) with radii of 20, 30, 34, and 50 degrees are shown in Fig.~\ref{f:sbub_image}. The wedges subtend the range of azimuthal angles between $\sim$270 and $\sim$360 degrees (counted clockwise from the Northern Galactic Pole).

\begin{figure*}
\centering
\includegraphics[angle=0,trim=1cm 5.5cm 1cm 2.5cm,width=0.95\columnwidth]{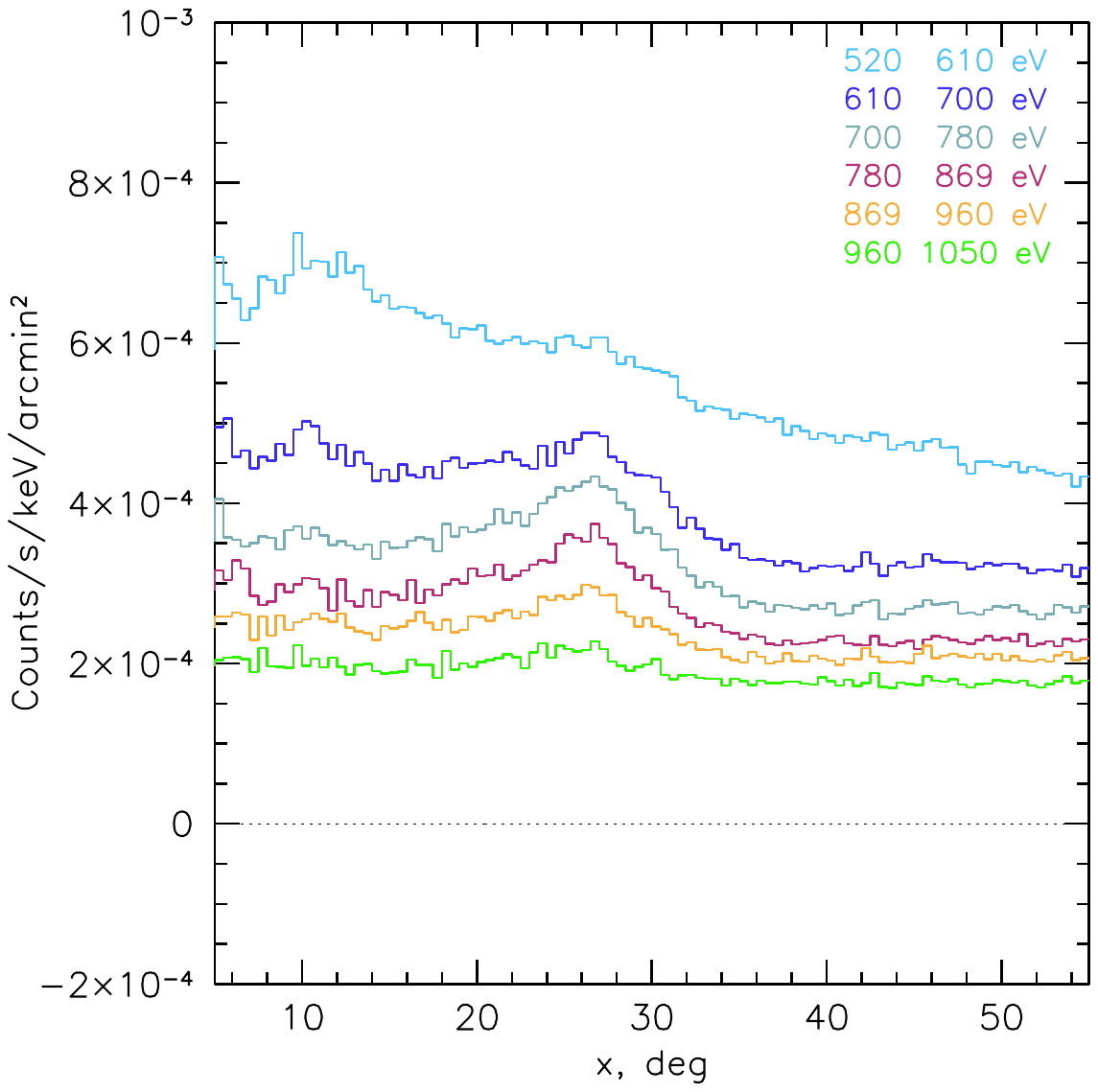}
\includegraphics[angle=0,trim=1cm 5.5cm 1cm 2.5cm,width=0.95\columnwidth]{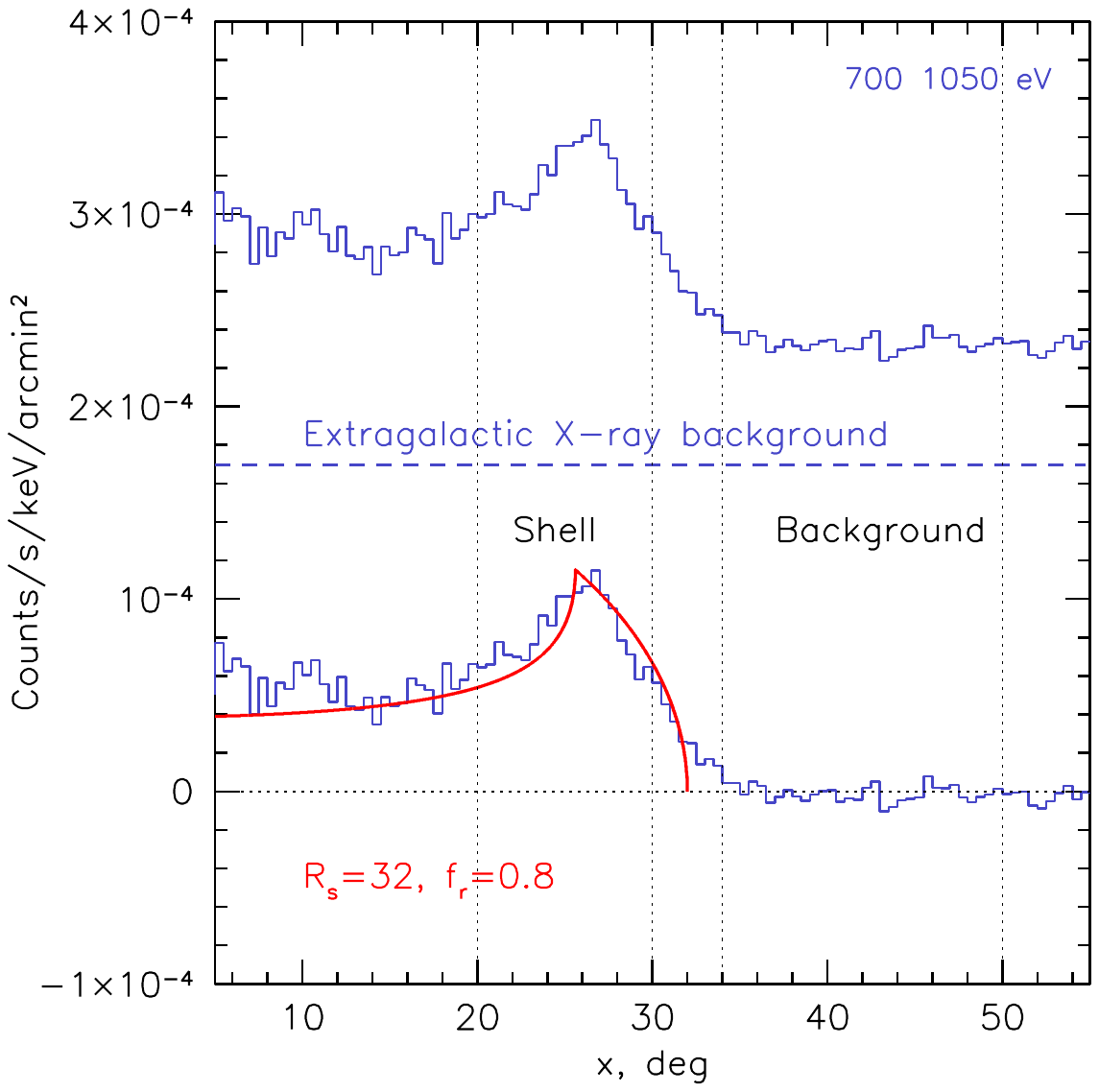}
\caption{Radial profiles across the Southern bubble in several energy bands. Statistical uncertainties associated with photon counting noise (at the level of 1-3 percent of the flux) are not shown.  
The data are accumulated in the same wedge as shown in Fig.~\ref{f:sbub_image} and using narrow radial bins. The left panel shows the observed profiles in narrow bands (the instrumental background is removed). The softest band (520-610~eV), dominated by the OVII line, to which the widespread diffuse Galaxy emission makes an important contribution, shows a largely linear trend with radii. On the contrary, the harder bands, which are presumably dominated by the lines characteristic of hotter plasma (e.g., O~VIII, Fe~XVII, Ne~IX, and Ne~X), all show a non-monotonic behavior with a clear bump peaking at 26-28 degrees from the adopted center.
The right panel shows the surface brightness profile in a broader band of 0.7-1.05~keV. The top blue histogram shows the observed profile, including the contribution of CXB (dashed horizontal line) and that of the Galaxy. The bottom histogram shows the same profile after subtraction of the background level, determined from the regions outside the Bubble. The peak in this band is $I_X\sim 3.5\times 10^{-5}\, {\rm counts\,s^{-1}\,arcmin^{-2}}$. For comparison, the red curve shows the expected surface brightness of a uniform spherical shell with the outer radius of $R_s=32$ degrees and the inner radius $0.8R_s$. The normalization of the curve was set to approximately match the observations.
Two pairs of vertical dotted lines show radial ranges used for spectra extraction.} 
\label{f:rprof}
\end{figure*}

The radial profiles, corresponding to such concentric wedges, of the X-ray emission in several energy bands are shown in the left panel of Fig.~\ref{f:rprof}. These bands are chosen to emphasize the contributions of ions such as O~VII, O~VIII, Fe~XVII (2 bands), Ne~IX, and Ne~X. The intrinsic detector background is subtracted. The lowest energy band (520-610~eV), containing the O~VII $K_\alpha$ complex, shows a steady, almost linear negative gradient with radius. In contrast, the brightness in all harder bands shows a prominent "bump" corresponding to the assumed shell position, followed by a spatially flat region at radii larger than $\sim 34$ degrees.   

The right panel of the same figure shows the radial profile in a broader energy range (700-1050~eV) before and after subtracting the sky background level determined from the radial range beyond $34$ degrees. For comparison, the red line shows the expected surface brightness profile for a uniform shell with the outer radius of 32 degrees and the inner radius $\sim0.8$ of the former. The consistency of the red line with the observed profile shows that, at least morphologically, the shell model is a reasonable approximation of the SE bubble. 

\subsection{Geometrical parameters of the shell}
\label{s:shell_geo}

To proceed with a more detailed model of the shell, below we summarize the main geometrical parameters of the shell adopted here, approximating the shell in the studied wedge as a part of a spherical layer.  

The angular size of the outer radius is $\theta\approx 32$ degrees. Therefore, the physical size (radius) is related to the distance of the sphere as $R_{\rm out}=D\sin\theta\approx 0.53D$. The angular size of the inner radius of the shell is $\approx 0.8\theta$. These two angles define the maximum size of the region along the line of sight that crosses the shell (near the inner radius) as $S=\eta D\approx 0.61 D$.

\section{Shell spectrum}
\label{s:spec}

\begin{figure}
\centering
\includegraphics[angle=0,trim=0cm 11.5cm 1cm 2.5cm,clip,width=0.95\columnwidth]{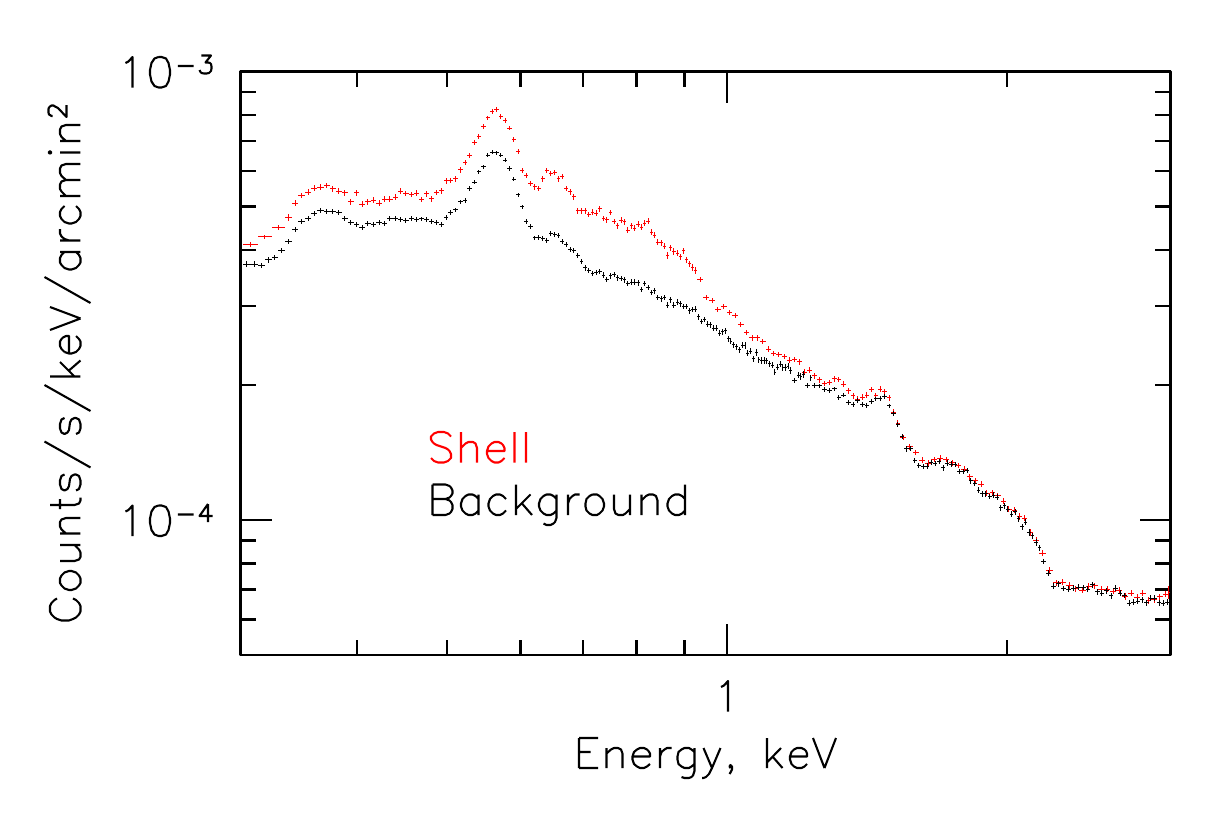}
\caption{Observed spectra of the bright shell and the background region (see the right panel in Fig.~\ref{f:rprof}). The spectra are normalized per unit solid angle (per square arcminute) and per one (out of seven) eROSITA telescope unit. The difference between these spectra is attributed to the shell emission.} 
\label{f:spec_2rings}
\end{figure}

Given that the X-ray morphology in the 0.7-1.05~keV band appears to be broadly consistent with the simple shell model, we proceed with the spectrum extraction. To this end, our default assumption is that the two radial bins shown in the middle panel of Fig.~\ref{f:sbub_image} with the green and blue colors correspond to the "source" and "background" regions, respectively. However, we now want to get the spectrum covering a broader band, down to 0.3~keV, to get the best constraints on the spectral parameters. Given that the solid angles subtended by both regions are substantial and there are significant variations in the Galactic absorbing column density, the difference between the source and background regions' spectra might be affected, especially at low energies. To avoid this problem, only regions where the total absorbing column density falls into a narrow range $N_{\rm H}=10^{19}-2.5\times 10^{20}\,{\rm cm^{-2}}$ were used for spectra collection. This results in a reduction of sky areas for the source and background regions, which are shown in the right panel of Fig.~\ref{f:sbub_image}. Additionally, bright compact sources and galaxy clusters have been uniformly detected and masked out (white circles in the green and blue areas). The resulting spectra (normalized per unit solid angle, namely per ${\rm arcmin^2}$) are shown in Fig.~\ref{f:spec_2rings}. As expected, there is a clear excess of the shell spectrum over the background spectrum at energies below $\sim 1.5\,{\rm keV}$.

\begin{figure*}
\sidecaption
\includegraphics[angle=0,trim=1cm 5.5cm 1cm 2.5cm,width=6cm]{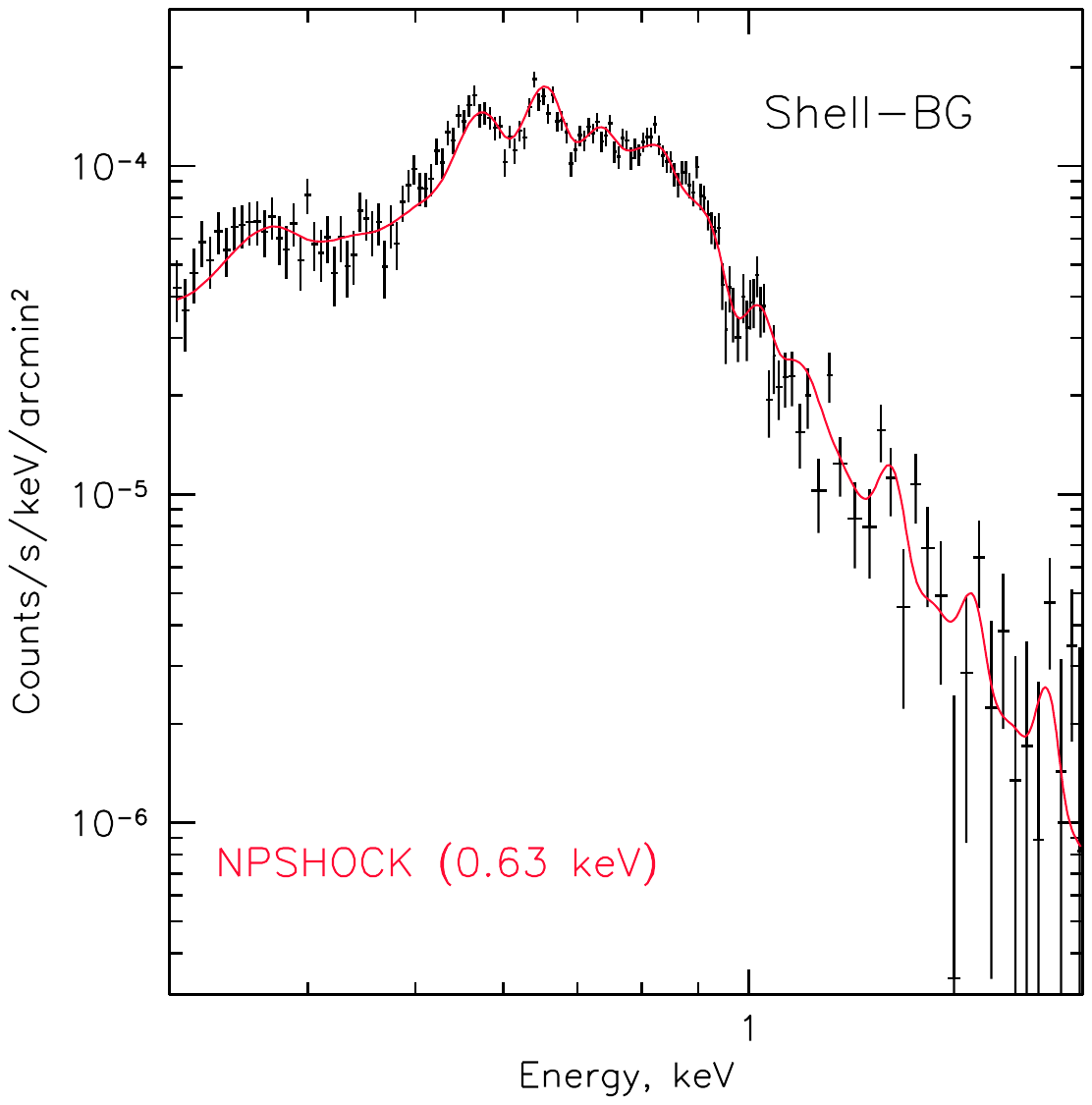}
\includegraphics[angle=0,trim=1cm 5.5cm 1cm 2.5cm,width=6cm]{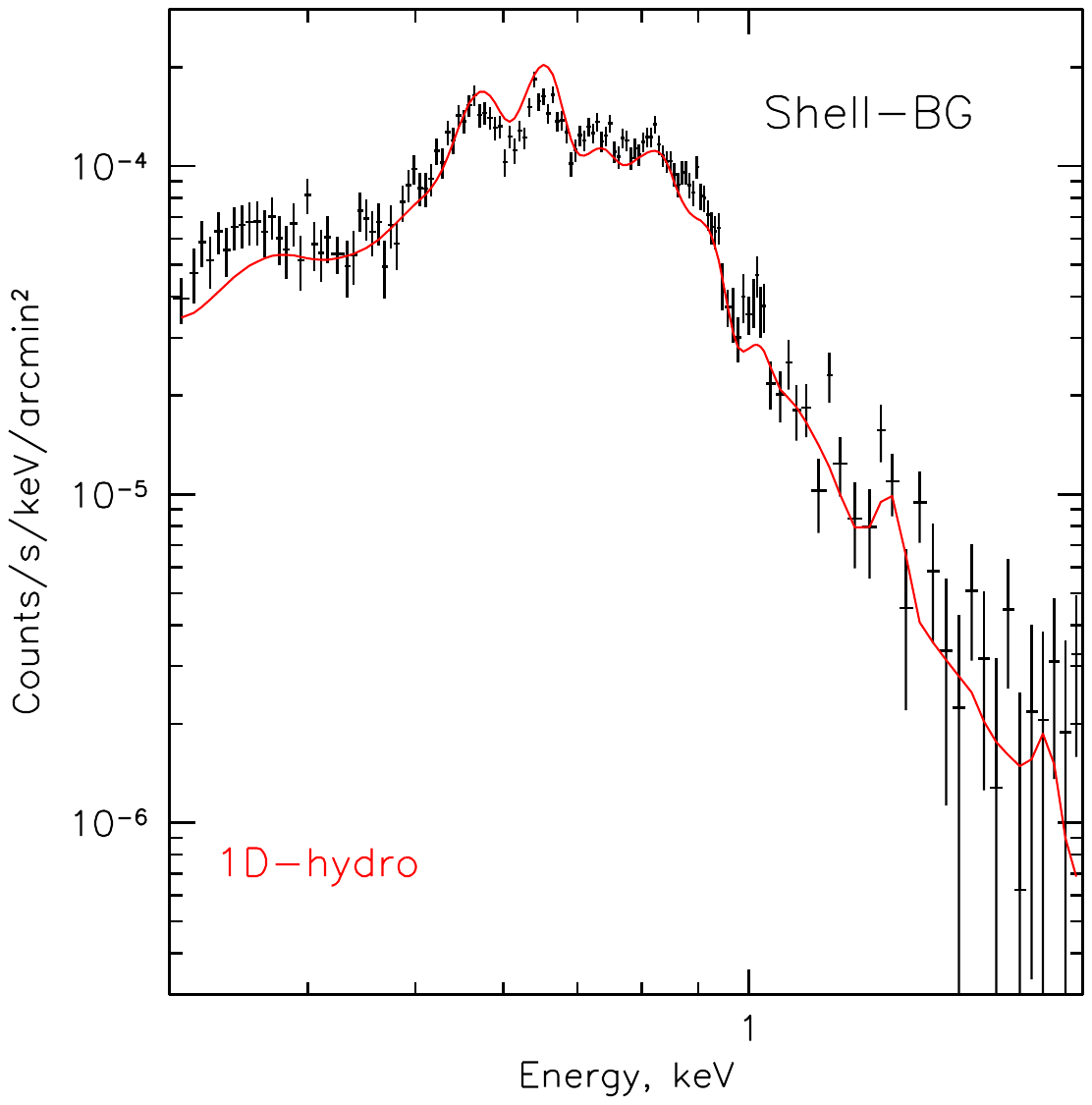}
\caption{Background-subtracted spectrum of the bright shell region (see the right panel in Fig.~\ref{f:sbub_image} for the definition of the spectra extraction regions). {\bf Left:} Comparison of the spectrum with the best-fitting \texttt{NPSHOCK} with final temperature $0.63\,{\rm keV}$ and the inonization parameter $\tau\sim 2\times 10^{11}\,{\rm cm^{-3}s}$. This model performs significantly better than other simple models like \texttt{APEC}. {\bf Right:} Comparison of the same spectrum with the predictions of the 1D hydrodynamic model. The red curve shows the predicted spectrum in the shell between $0.7R_s$ and $R_s$ for the \texttt{C2L6e41} model, with 
 the downstream temperature and density motivated by the spectral analysis of the observed spectrum. The best-fitting normalization (shown in the plot) is a factor $\sim 1.6$ higher than the initial model.}
\label{f:spec_npshock}
\end{figure*}

The "net" shell spectrum (the difference between the source and background spectra) is shown in Fig.~\ref{f:spec_npshock}. Several prominent emission lines are clearly present in the spectrum, including those of O~VII, O~VIII, Fe~XVII, and weaker lines of Ne~IX and Ne~X. These lines are characteristic of different temperatures for plasma in ionization equilibrium. Indeed, the \texttt{APEC} model \citep{2001ApJ...556L..91S} provides a poor fit to the spectrum (see Table~\ref{t:models}), emphasizing the need for multi-temperature gas or a departure from the CIE. One can describe the observed spectrum with a combination of several equilibrium components. However, a more attractive option is to use a propagating shock model. Indeed, \texttt{NPSHOCK} model \citep{2001ApJ...548..820B} provides a much better fit to the spectrum (see Table~\ref{t:models} and Fig.~\ref{f:spec_npshock}). 

The best-fitting abundance of heavy elements is rather low, $\sim 0.1$~Solar (when the elemental abundances from \citealt{1989GeCoA..53..197A} are used). Experiments with changing the abundance of iron relative to other elements show that Fe/O is consistent with the solar value, with a marginal preference for the Fe overabundance by a factor of $\sim 1.5$. Models with Fe/O ratio significantly lower than 1 provide a poor fit to the data.

Given that the \texttt{NPSHOCK} model provides a reasonable description of the shell spectrum, one can try to build a self-consistent propagating shock model, which we describe in the next section. We note here that for our setup, the \texttt{NPSHOCK} model is not fully adequate because the initial ionization state in this model corresponds to a neutral medium, while in the real CGM, elements like C, N, and O might be fully or partly ionized. We, therefore, use  \texttt{NPSHOCK} to identify a plausible parameter range and then use our NEI model to account for this issue.

\begin{table}[]
\caption{Models approximating the background-subtracted shell spectrum in the 0.3-2~keV band (235 spectral bins).}
    \centering
    \begin{tabular}{l|l|l}
    \hline
    \hline
    Model & Parameters & $\chi^2$\\
    \hline
    \hline
        \texttt{APEC} & $kT=0.295 \pm  0.0026 \,{\rm keV}$ & 740.6 \\
        & $A=0.17 \pm 0.012$ & \\
         \hline
        \texttt{NPSHOCK} & $kT_a= 0.64 \pm  2.2\times 10^{-2} \,{\rm keV}$ & 268.6 \\
       &  $kT_b= 0.15 \,{\rm keV; frozen}$ & \\
       & $\tau=2.8\times 10^{11}  \pm 1.6\times 10^{10}\,{\rm cm^{-3}s}$ & \\
        & $A=0.10 \pm 0.008$ & \\   
        \hline
        \texttt{C2L6e41}  & $R=\left [0.7R_s : R_s\right ]$; frozen & \\
        Steady wind & $A=0.1$; frozen & \\
        & Normalization; frozen ($1.0$)& 2297.2 \\
        & Normalization; free ($1.59$)  & 482.3 \\
          \hline
        \texttt{C2L6e42}  & $R=\left [0.7R_s : R_s\right ]$; frozen & \\
        $\sim$Sedov& $A=0.1$; frozen & \\
        & Normalization; frozen ($1.0$)&  4368.0\\
        & Normalization; free ($2.1$)& 782.9 \\
          \hline
       \hline
    \end{tabular}
    \tablefoot{
    A single-temperature \texttt{APEC} (the abundances are taken from \citealt{1989GeCoA..53..197A}) provides a poor fit to the spectrum. A much better approximation is provided by the \texttt{NPSHOCK} model. In this model, the initial electron temperature ($kT_b$) is poorly constrained, and we fix it at $0.15\,{\rm keV}$. Varying $kT_b$ between 0 and 0.2~keV does not change $\chi^2$ by more than $1$.
    The next model, \texttt{C2L6e41}, formally has no free parameters; the information on the geometry of the shock, the shock velocity, and the upstream gas density in the hydrodynamic model comes from the analysis of the observed spectrum. The model performs reasonably well, provided its normalization is set free, which requires a factor $\sim 1.6$ higher flux and translates to increasing the upstream density by $\sqrt{1.6}\approx 1.26$ (formally, other parameters have to be adjusted, too). Another possibility to boost the X-ray flux by the same factor is via increasing the abundance of metals ( e.g., going from $A=0.1$ to $A\approx 0.18$ would change the 0.7-1.05~keV flux by a factor of $\sim 1.6$, see Appendix~\ref{a:abund}). This suggests the shell spectrum can indeed be described as a propagating shock model. The last model, \texttt{C2L6e42}, is similar to \texttt{C2L6e42} but has an order of magnitude more powerful wind that runs only for a short time $\sim 0.5\,{\rm Myr}$. The rest of the evolution is more similar to the Sedov-Taylor case and 
    leads to a lower ionization parameter in the shocked gas (see Appendix~\ref{a:ib_spec}). Clearly, this model performs worse than \texttt{C2L6e41} even if the normalization is treated as a free parameter. However, one can not exclude that a more elaborate model can perform better.}
    \label{t:models}
\end{table}

\section{1D model and simplified geometry}
\subsection{Constraints on the size/distance}

From images and spectra, we can get four major observables related to the bright shell. Namely, its angular size $\theta_{\rm obs}$, characteristic surface brightness $I_{X,\rm obs}$, the gas temperature $T_{\rm obs}$, and the ionization parameter $\tau_{\rm obs}=n_{e}\cdot t$. Two additional observables (in the context of the \texttt{NPSHOCK} model), the abundance of heavy elements $A_{\rm obs}=Z/Z_\odot$ and the initial temperature $T_{0,\rm obs}$, are treated as "secondary" and used below for the consistency checks. In the reference model, the abundances from \cite{1989GeCoA..53..197A} are used.

Our goal is to obtain the simplest possible physical model that can reproduce these observables. To minimize the number of free parameters, we adopted a model of a spherically symmetric shock in a uniform medium driven by a steady fast wind. This model has four major parameters: the distance $D_{\rm m}$ to the shell, the downstream gas density $n_{e,\rm m}$, the mechanical power of the winds $L_{\rm m}$, and the abundance of heavy elements in the gas $A_{\rm m}=Z/Z_\odot$. This model also has two "secondary" parameters, namely, the velocity of the wind $\varv_{\rm m}$ and the initial gas temperature $T_{0,\rm m}$.

The logical sequence of deriving constraints on the model parameters from the observables could be outlined as follows:
\begin{itemize}
\item Given $D_{\rm m}$ fixes the size of the shell $R_{\rm m}=D_{\rm m}\sin \theta_{\rm obs} \approx D_{\rm m}\theta_{\rm obs}$ (see Sec.~\ref{s:shell_geo}).
\item The observed gas temperature $T_{\rm obs}$ (see Sec.~\ref{s:spec}) fixes the shock velocity $\varv_{s,\rm m}$ (under the assumption that the shock is strong), i.e. 
$T_{\rm obs}=(3/16) \mu  m_p \varv_{s,\rm m}^2$, where $\mu$ is the mean atomic weight and $m_p$ is the proton mass. Together with $R_{\rm m}$, this sets the characteristic age of the structure, e.g., $t_{\rm m} =\zeta R_{\rm m}/\varv_{s,m}$ as a function of $D_{\rm m}$. Here, the prefactor $\zeta\sim O(1)$ characterizes the expansion regime, with $\zeta=2/5$ for the Sedov-Taylor strong explosion problem and $\zeta=3/5$ for the wind case (assuming that the shock is strong).

The gas density is then derived from the observed ionization parameter $n_{e,\rm m}=\tau_{\rm obs}/t_{\rm m}$ as a function of $D_{\rm m}$. Namely,
\begin{equation}
    n_{e,{\rm m}}=\frac{\tau_{\rm obs}}{\zeta D_{\rm m}\theta_{\rm obs}}\left ( \frac{T_{\rm obs}}{(3/16) \mu  m_p} \right)^{1/2}
    \label{e:ne-tau}
\end{equation}

\item Knowing $I_{X,{\rm obs}}$ (in a particular energy band), one can place an additional constraint on the distance using
\begin{equation}
I_{X,\rm obs} \sim n_{e,\rm m}^2 \eta R_{\rm m} \varepsilon(T_{\rm obs},\tau_{\rm obs},A_{\rm m})\propto \varepsilon(T_{\rm obs},\tau_{\rm obs},A_{\rm m})/D_{\rm m},
\label{e:ixobs}
\end{equation}
\end{itemize}
where the dimensionless constant $\eta\sim 1$ (see Sec.~\ref{s:shell_geo}) relates the length of the line-of-sight in the hot gas with the shell radius (see Sec.~\ref{s:shell_geo}), and $\varepsilon(T_{\rm obs},\tau_{\rm obs},A_{\rm m})$ is the gas emissivity in this energy band. Should $\varepsilon$ scale linearly with abundance, the r.h.s. of eq.~\ref{e:ixobs} would be $\propto \frac{A_{\rm m}}{D_{\rm m}}$, leading to degeneracy between distance and abundance, with smaller distances requiring lower abundances. However, at low abundances, the linear dependence breaks (see Appendix~\ref{a:abund}), imposing a lower limit on the distance.

\begin{figure}
\centering
\includegraphics[angle=0,trim=1cm 5.5cm 1cm 2.5cm,width=0.9\columnwidth]{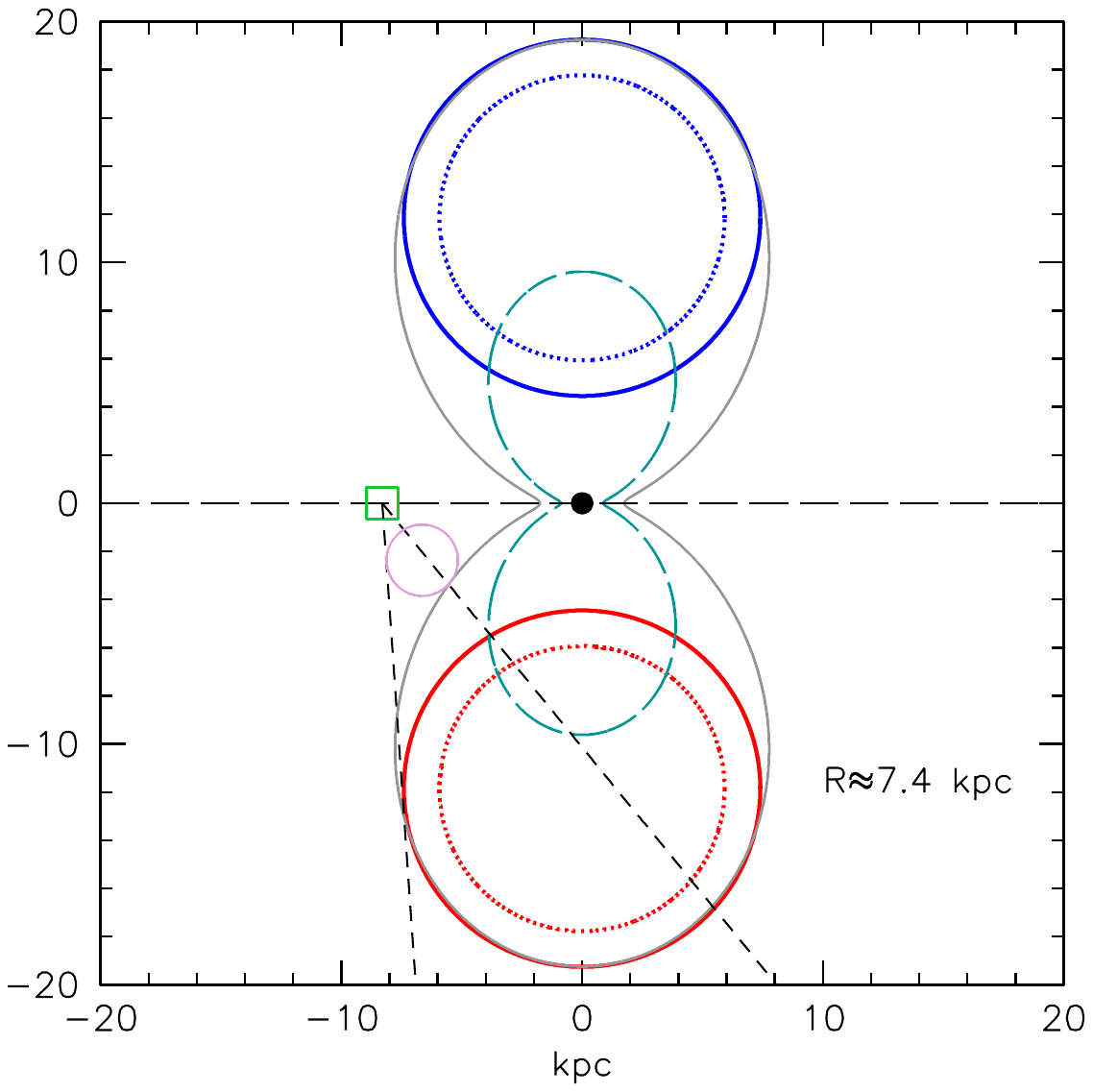}
\caption{Sketch of the Southern eROSITA Bubble based on the morphological and spectral analysis. The observer's position is shown with a green box. The Southern Bubble visible outer boundary is approximated by a sphere/shell (red lines) with a radius of $\sim 7.4\,{\rm kpc}$. This sphere is plausibly a part of a more complicated structure sketched by the gray lines.
The observed boundary of the bubble is at a distance of $\sim 12\,{\rm kpc}$.  The outer boundary of the Fermi bubbles is shown with a green line.  The blue lines show a simple reflection of the SB model to the North.}
\label{fig:sketch2}
\end{figure}

\begin{figure}
\centering
\includegraphics[angle=0,trim=1cm 5.5cm 1cm 2.5cm,width=0.9\columnwidth]{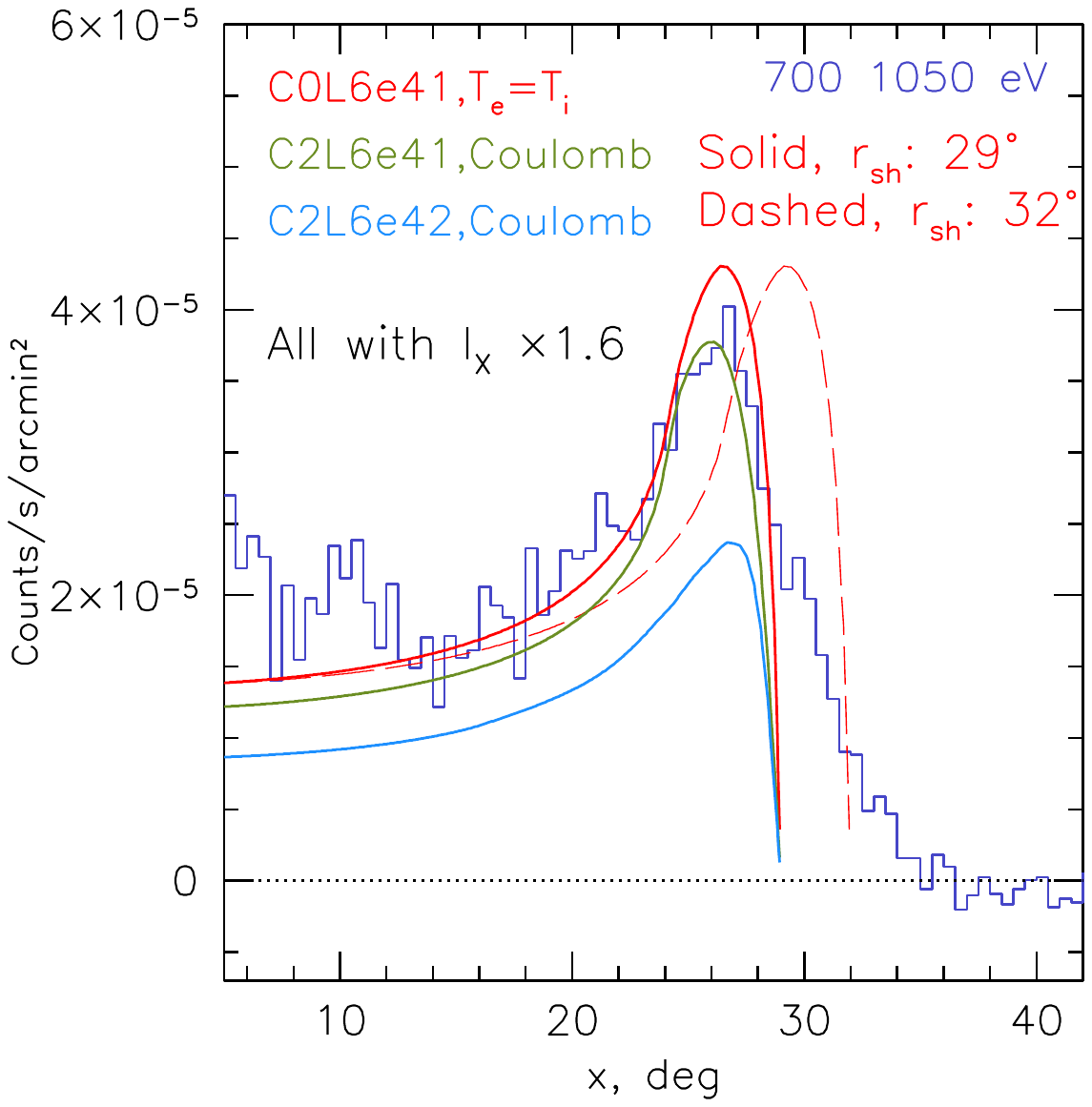}
\caption{Comparison of the observed radial profile with the predictions of the 1D model. Two solid curves illustrate the impact of electron-ion temperature equilibration: the red curve shows the $T_i=T_e$ case, while the blue curve is for the equilibration mediated by Coulomb scattering. In both cases, the position of the forward shock in the model is at $29^\circ$ with a steady energy release rate of $L=6\times 10^{41}\,{\rm erg\,s^{-1}}$. The red dashed line shows the case when the position of the forward shock is at $32^\circ$ from the center.
The brown solid line shows the model for $L=6\times 10^{42}\,{\rm erg\,s^{-1}}$ for a short time, which is similar to the Sedov-Taylor problem. All curves predicted by the fiducial model have been multiplied by a factor of 1.6 to better match the observed profile. The model has a sharper "edge" than the data, which is plausibly the shortcoming of our simplistic geometrical model that approximates the edge with a circle. Much deeper images are needed to determine the shape of the X-ray shell unambiguously. }
\label{f:rmodels}
\end{figure}

In practice, we slightly modified the scheme described above. We fixed geometrical parameters (angular sizes) according to Sec.~\ref{s:shell_geo} and $T_{\rm obs}=0.64\,{\rm keV}$, used Eq.~\ref{e:ne-tau} to express $\tau$ through $n_e$, and solved Eq.~\ref{e:ixobs} for three values of metal abundance, $A_{m}=0.05, 0.1, 1$. This procedure yields the values of $n_{e,{\rm m}}$ and $\tau_{\rm m}$ as a function of the emitting region size (along the l.o.s.). The value of $\tau_{\rm obs}$ is then used to select the preferred model as shown in Fig.~\ref{f:ix_size}. This procedure unambiguously selects large values of $S\sim 5-10\,{\rm kpc}$ and "local" solutions with $S\lesssim 1 \,{\rm kpc}$ are excluded. This conclusion is robust against any uncertainties/assumptions made when solving for $n_e$ and $\tau$. 

Given that in the model the size $S$ and the distance $D$ are related (see Sec.~\ref{s:shell_geo}) and the downstream density can be estimated from  Fig.~\ref{f:ix_size}, we can select a set of parameters that can be used to characterize the shell. When doing so, we round some numbers to emphasize that the model is not intended to reproduce observational results precisely. Rather, we would like to verify if there are any major inconsistencies. The parameters of the model are summarized in Tab.~\ref{t:params}.

\begin{table*}[]
\caption{Adopted model parameters for 1D hydrodynamic model.}
    \centering
    \begin{tabular}{l|l|l}
    \hline
    \hline
    Parameter & Value & Comment \\
    \hline
    \hline
        Shock radius & $R_{\rm m}\simeq8\,{\rm kpc}$ & from Fig.~\ref{f:ix_size} \\
        Upstream density & $\rho_0\simeq3\times 10^{-4} \,{\rm m_p\,cm^{-3}}$ & from Fig.~\ref{f:ix_size} assuming compression factor $C\sim3.1$  \\
            Upstream temperature & $T_{0,{\rm m}}=0.15 \,{\rm keV}$ & Assumed as a plausible value for the MW CGM; see also Fig.~\ref{f:kt0}  \\
       Metal Abundance & $A_{\rm m}\simeq 0.1$ &  from Fig.~\ref{f:ix_size}; relative to \cite{1989GeCoA..53..197A}\\
       Wind velocity & $\varv_{\rm w}= 4\times 10^{4}\,{\rm km\,s^{-1}} $ &  Large value $\Rightarrow$ small radius of the termination shock \\
       Wind power & $L= 6\times 10^{41}\,{\rm erg\,s^{-1}}$; $t_{\rm age}\approx 7.5\,{\rm Myr}$ &  \texttt{C2L6e41}; Continuous   \\
     Wind power & $L= 6\times 10^{42}\,{\rm erg\,s^{-1}}$; $t_{\rm age}\approx 5.1\,{\rm Myr}$ & \texttt{C2L6e42}; Short; $\times 10$ more powerful wind operating for $0.5\,{\rm Myr}$ \\
          \hline
       \hline
    \end{tabular}
    \tablefoot{The age of the system $t_{\rm age}$ and the ionization parameter $\tau=\int_0^t n_e(t) dt$ are evaluated at the moment when the shock reaches $R_{\rm m}$.}
    \label{t:params}
\end{table*}

For the above set of parameters, we run a simple 1D pure hydrodynamic model using the \texttt{PLUTO} code \citep{2007ApJS..170..228M}. A compact source\footnote{This 1D model ignores density gradients in the ambient medium and the offset of the bubbles' centroid from the Galactic Center; see Fig.~\ref{f:sbub_image}} 
produces an isotropic steady wind with total kinetic power $L$. The wind velocity is set to a large value to ensure that the wind termination shock is very close to the center, well inside the contact discontinuity (CD) between the shocked wind and the shocked CGM. The wind expands into a homogeneous medium (see Tab.~\ref{t:params}), and the simulations stop when the forward shock reaches the radius $R_{\rm m}$. For our baseline model, this occurs at time $t \sim 7.5\,{\rm Myr}$.

We then run an additional simulation with a factor of 10 more powerful wind, which is launched for $\sim 0.5\,{\rm Myr}$. The duration was chosen so that the total energy release is comparable (within a factor of 2), and the forward shock sonic Mach number at $R_{\rm m}$ is approximately the same. In this run, the shock reaches $R_{\rm m}=8\,{\rm kpc}$ in $\sim 5.1\,{\rm Myr}$. 

These simulations yield radial profiles of the gas parameters. For each Lagrangian fluid element in the profile, we know the time evolution of its density and temperature after this element has passed through the shock. These data are used to calculate the effective ionization time $\tau$ for individual cells. The corresponding profiles are shown in Fig.~\ref{f:hydro_model}. With our choice of the fast wind velocity $\varv_{\rm w}= 4\times 10^{4}\,{\rm km\,s^{-1}} $, the density of the shocked wind is always much lower than the density of the shocked CGM. This means that only the gas layers beyond CD can contribute to the X-ray emission. Appendix~\ref{a:ib_spec} provides some basic details on the generation of projected spectra based on the hydrodynamic simulations. These spectra can be directly compared with the observational data as illustrated in the right panel of Fig.~\ref{f:spec_npshock}.  Despite relative simplicity, the \texttt{CL2L641} model provides a reasonable description of the spectrum. We note here that the data are compared with the model prediction at a given projection radius, where the surface brightness peaks, rather than the annulus-averaged model. We concluded 
that not only the morphology of the shell, but its spectrum, too, is broadly consistent with the propagating shock scenario. On the other hand, the   \texttt{CL2L642} model clearly performs worse, indicating that the data can (at least statistically) differentiate between these models.

The predictions of the models can be further directly compared with the X-ray surface brightness profiles. This is done in Fig.~\ref{f:rmodels}. The models broadly reproduce the shell geometry and demonstrate the moderate (but not negligible) impact of the additional assumptions, such as the timing of the energy release or the role of Coulomb energy exchange.  The parameters of the models could be further tuned to reproduce the data even better, but there is no guarantee that these models will be closer to the real properties of the observed structures.

\begin{table*}[]
\caption{Variations around the baseline model in comparison with observed spectra.}
    \centering
    \begin{tabular}{llllll}
    \hline
    \hline
    Model & Wind power $L$ & Upstream density $\rho_0$ & Abundance & $N$ & $\chi^2$ (235 bins)\\
    & ${\rm erg\,s^{-1}}$ & ${\rm m_p\,cm^{-3}}$ &  & & \\
    \hline
    \hline
        Baseline wind model & $ 6.0\times 10^{41}$& $3.0\times 10^{-4} $ & $A_{\rm m}=0.1$ & $1.58$ &  $482.3$\\
        Higher abundance & $6.0\times 10^{41}$& $3.0\times 10^{-4}$ & $A_{\rm m}=0.2$ & $0.87$ &  $664.3$\\
        Higher power and density & $ 1.2\times 10^{42}$ & $6.0\times 10^{-4} $ & $A_{\rm m}=0.1$ & $0.46$ &  $390.8$\\
    Higher power, density, abundance & $1.2\times 10^{42}$ &$6.0\times 10^{-4}$ & $A_{\rm m}=0.3$ & $0.18$ &  $673.5$\\
   Lower power and density & $ 3.0\times 10^{41}$ & $1.5\times 10^{-4}$ & $A_{\rm m}=0.1$ & $5.84$ &  $1098.5$\\
     Lower power, density, higher abundance & $3.0\times 10^{41}$ & $1.5\times 10^{-4} $ & $A_{\rm m}=0.3$ & $2.15$ &  $1395.4$\\
       Colder medium (0.015 keV), higher power & $1.2\times 10^{42}$& $3.0\times 10^{-4} $ & $A_{\rm m}=0.1$ & $1.30$ &  $446.9$\\   
          \hline
       \hline
    \end{tabular}
    \tablefoot{ The last two columns show the best-fitting normalization $N$ of this model and the value of $\chi^2$, corresponding to it. A "better"-performing model should have $N$  closer to $1$ and a lower $\chi^2$. The metal abundances are relative to \citealt{1989GeCoA..53..197A}.
    Changing the wind power and the gas density by the same factor preserves the age of the system and the temperature downstream of the shock for a fixed shock radius. For the "Colder medium, higher power" case, the upstream density was unchanged, while the wind power was increased by a factor of 2 to increase the Mach number of the shock so that the downstream temperature is in the right range. From this analysis, we concluded that the density and power are estimated with a factor of $\sim 2$ uncertainty. The constraints on the minimal upstream temperature are weak.}
    \label{t:modelvariations}
\end{table*}

\section{Discussion}
\label{s:discussion}
\subsection{Milky Way CGM}
\label{s:cgm}
\begin{figure}
\centering
\includegraphics[angle=0,trim=1cm 5.5cm 1cm 2.5cm,width=0.9\columnwidth]{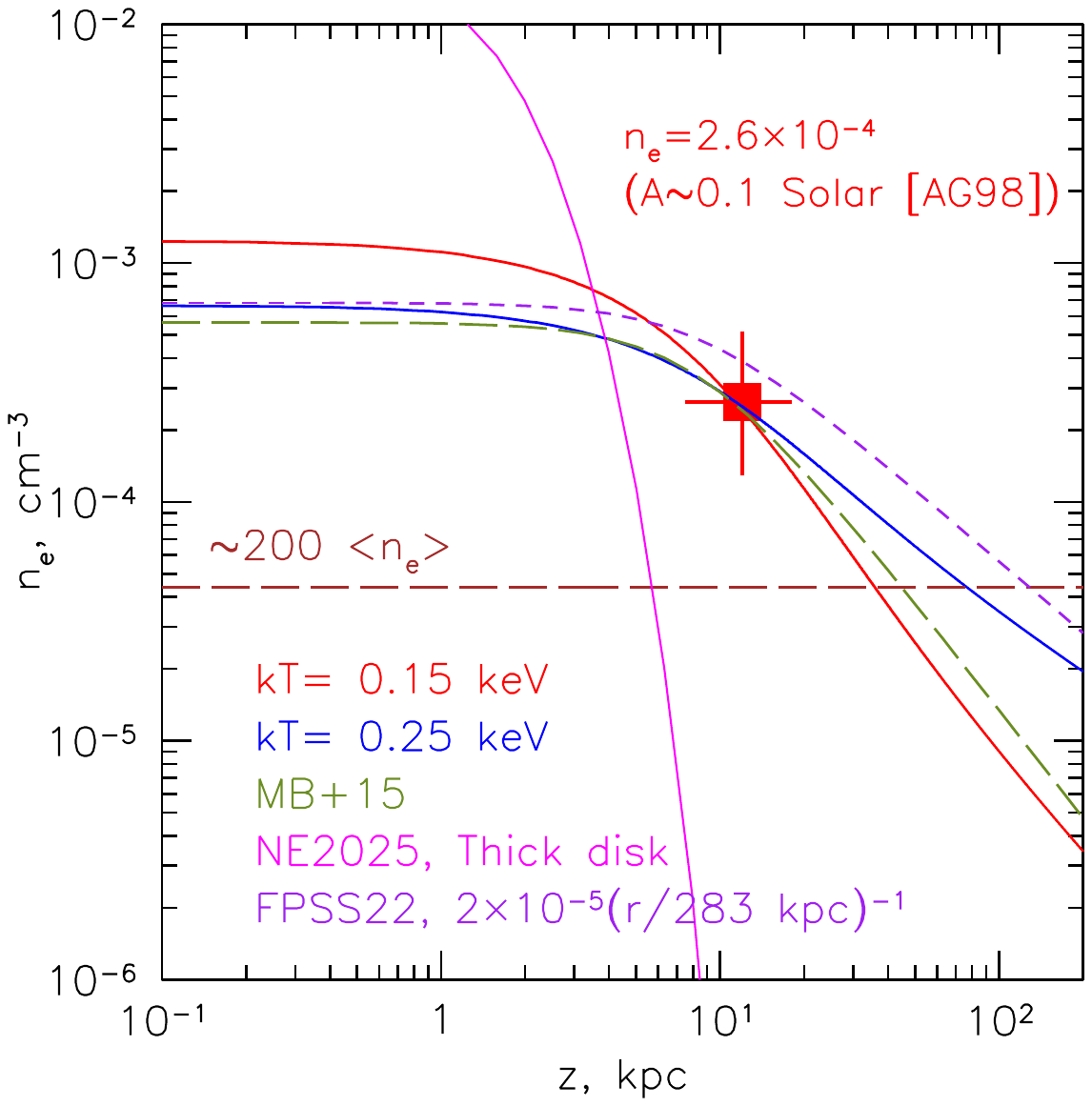}
\caption{Comparison of the density measurement provided by SE bubble spectral analysis (red box) with other data (as a function of the distance $z$ above the disk near the Sun position). The (reasonably conservative) error bars associated with the SE bubble measurement include i) in the horizontal direction, the range of $z$ contribution to the shell in the fiducial model, and ii) a factor of 2 in density that encapsulates the uncertainties in geometry and abundance. 
The magenta line shows the thick disk model derived from the dispersion measurements \citep{2026ApJ..1002....3O}. The green dashed line shows the expected $n_e(z)$  dependence in the radial model of \cite{2015ApJ...800...14M}, while the purple line shows the "standard" model of \cite{2022ApJ...928...37F}. These two models match the SE-bubble point well, although the abundance of metals adopted in our model ($\sim 0.1$ relative to \citealt{1989GeCoA..53..197A}) is lower. The blue and red lines show the $z$ dependence of the density for hydrostatic atmospheres, normalized at the SE bubble point, in the Milky Way potential with $kT=0.25$ and $0.15\,{\rm keV}$, respectively.} 
\label{f:mw_zprof}
\end{figure}

Taking at face value, the X-ray data provide constraints on the CGM properties: (i) the gas density is $\rho\sim 3\times 10^{-4}\,m_p{\rm cm^{-3}}$ (equivalent to $n_e\sim 2.6\times 10^{-4}\,{\rm cm^{-3}}$ for He/H number density ratio of 1/12) at a distance of $R\sim 12\,{\rm kpc}$ from the Sun and at a similar hight $z\sim 12\,{\rm kpc}$ above the disk.  This value is shown in Fig.~\ref{f:mw_zprof} as a red box. The shown error bars are dominated by systematic rather than statistical uncertainties. The horizontal error bars reflect the range of heights above the disk that, in our model, contribute to the observed emission. The vertical error bars (factor of 2) include the uncertainties in the assumed geometry and abundance. Since the density scales as the square root of size or the metallicity pre-factor in emissivity, we consider this choice of the plotted error bars as conservative.  

We supplement this qualitative statement with a few more quantitative tests. Namely, we run several models analogous to the \texttt{C2L6e41} model, but vary the ambient gas density and the abundance of heavy elements. In doing so, we change the density and the power of the wind by the same factor. As a result, the size and shock velocity, and, therefore, the downstream plasma temperature, are unchanged. Only the volume emissivity and/or the ionization parameter change. The resulting models are compared with the observed spectrum, as shown in Table~\ref{t:modelvariations}. Overall, these experiments show that our estimates of density and abundance are subject to uncertainties of a factor of $\sim 2$.  Given this uncertainty, one can ignore factors of order unity related to the presence of He or other species, round the result to one digit,  and conclude that $n_p\approx n_e\approx 3\times 10^{-4}\,{\rm cm^{-3}}$. 

As mentioned in Sect.~\ref{s:spec}, the \texttt{NPSHOCK} model in XSPEC assumes that the upstream gas is neutral. We (approximately) reproduce this case by setting the initial temperature in our model ten times lower than in the baseline model, and still get a qualitatively reasonable approximation of the observed spectrum. This implies that the constraints on the initial (upstream) temperature are poor within the frame of the \texttt{NPSHOCK} model. The information on the initial temperature is related primarily to the low-energy part of the observed spectrum, e.g., lines of C, N, and, in particular, of O~VII. We further illustrate the role of the upstream temperature on the model spectra in Appendix~\ref{a:kt0} (see Fig.~\ref{f:kt0}). Formally, the changes in the ratio between the OVII and OVIII lines can tightly constrain the upstream temperature. However, the OVII~$\rm K_\alpha$ line
is strong in the "foreground" emission (see Fig.\ref{f:rprof}), and it is difficult to remove it (and other low-energy emissions) cleanly for a large region like the one used here. In addition, the spatially variable absorption across the source and background regions might play a role. While we tried to minimize these effects, some bias might still be present. We therefore consider that we currently can not get very tight constraints on the upstream temperature $T_0$, although it is in principle possible with the available statistics. Instead, we assume that $kT_0$ is somewhere between 0.15 and 0.20 keV, which is in better agreement with the data than lower or higher values (see Fig.\ref{f:kt0}). For the baseline model, we set $kT_0=0.15\,\rm keV$.

In what follows, we assume that the gas is isothermal and is in hydrostatic equilibrium in the Milky Way potential. One can estimate the density profile perpendicular to the Galaxy disk at the Sun's position, shown in Fig.~\ref{f:mw_zprof}, using the measured electron density in the vicinity of the SE bubble to normalize the curves.  To this end, we use the approximation of the Milky Way potential from \cite{2016A&A...593A.108B} and assume that the gas is not rotating. 
The red and blue curves show the density distributions derived for 
$kT_{0,{\rm max}}=0.15$ and $0.25\,{\rm keV}$, respectively. 

It is interesting to compare these numbers with other models that use other methods to characterize the gas spatial density distributions in the Galaxy. The magenta, green, and purple lines show three such models. The first one (the magenta line) is the NE2025 model of \cite{2026ApJ..1002....3O} that uses radio sources and the dispersion measure to infer electron column density. Clearly, this model (the thick disk component of NE2025) is much more concentrated towards the Galactic disk and describes a component that becomes subdominant for $z\gtrsim 10\,{\rm kpc}$. The two other curves, from \cite{2015ApJ...800...14M} (dashed green) and \cite{2022ApJ...928...37F} (dashed purple), are in reasonable agreement with the SE Bubble measurement. We note in passing that the metallicity of the gas found in this study ($Z/Z_\odot \lesssim 0.1$) is lower than that derived in \cite{2022ApJ...928...37F} ($Z/Z_\odot\sim 0.3$). However, we used individual elemental abundances from \cite{1989GeCoA..53..197A}, which is a default set in the \texttt{APEC} family of spectral models, while in \cite{2022ApJ...928...37F} the abundances from \cite{2009ARA&A..47..481A} were adopted. In \cite{2009ARA&A..47..481A} and \cite{2003ApJ...591.1220L}, the abundances of key elements relevant for this study, namely, O, Ne, and Fe, are a factor of $\sim0.6-0.7$ lower than in \cite{1989GeCoA..53..197A}. As a result, using the abundances from  \cite{2009ARA&A..47..481A} results in the metallicity $Z/Z_\odot\sim 0.2$ for the same models and yields comparable fit statistics.

A comparison of the electron density distribution derived from the dispersion measurements \citep[e.g.][]{2026ApJ..1002....3O} with the hot and hydrostatic atmospheres (magenta vs red and blue lines in Fig.~\ref{f:mw_zprof}) shows that their profiles are very different, meaning that either the electrons that dominate the dispersion measure come from the cooler gas closer to the disk plane, or, alternatively, the medium is multiphase and there are pockets of hot gas in the cooler atmosphere. It is also plausible that the abundance of metals is higher at lower heights above the disk, so that from the Sun's position, this component makes a significant or even dominant contribution to the line absorption. For instance, \cite{2018MNRAS.474..696G} found that a flattened component with the scale-height $z_f\sim 0.36\,{\rm kpc}$ (in fact, $z_f$ is in the range between $0.14$ and $1.1\,{\rm kpc}$) contributes about a half of the absorbing column density for lines characteristic for $2\,{\rm MK}$ gas.

There were several studies of the X-ray emitting gas at various locations between the outer boundary of Fermi bubbles and the outer boundary of eROSITA bubbles \citep[e.g.,][]{2013ApJ...779...57K,2015ApJ...807...77K,2021ApJ...908...14K,2016ApJ...829....9M,2022MNRAS.512.2034Y}. In terms of spectral parameters, our results are most close to those of \cite{2022MNRAS.512.2034Y}, namely, the presence of NEI plasma in their analysis with the ionization parameter $\tau\sim 10^{11}{\rm cm^{-3}~s}$, gas temperature of up to 0.7~keV, and low abundance of metals $A\sim 0.2$. Their analysis was done for a set of small regions in the NPS and Loop I. One could use this agreement as an argument that the global properties of the Southern and Northern bubbles are qualitatively similar. Of course, line-of-sight projection of plasma with different temperatures might masquerade as NEI gas. However, for the "global" spectrum of the shell, the contributions of components with different ionization parameters should scale with $\tau$ (for a plane shock). The good agreement between the model and the observed spectrum makes the case for a propagating shock scenario stronger for the SE Bubble.

\subsection{More complicated models.}
Our 1D hydrodynamical model of an energy release in a uniform medium is, of course, a gross oversimplification. It ignores possible anisotropies in the central engine and ignores the density/temperature gradients that should be present in the gas.
However, the shock velocity comes essentially from the downstream temperature, while the density is related to the X-ray surface brightness and the ionization parameter estimates from the spectra. These are largely "local" (to the shell) quantities. Therefore, while the details of the central engine energy release might affect the total energy and the time-dependent behavior, the CGM properties at the visible edge of the SE bubble and the shock velocity are relatively robust. There are many more elaborate models, including full cosmological simulations \citep[e.g.,][and references therein]{2000ApJ...540..224S,2022MNRAS.514.2581M,2021MNRAS.508.4667P,2024MNRAS.535.1721P,2023ApJ...951...36S,2024A&ARv..32....1S,2026arXiv260200226G}, but for our purposes, the simplified model makes the connection to the underlying assumptions and associated uncertainties more transparent.

In Sect.~\ref{s:cgm}, we assigned a factor of 2 uncertainty to the measured density. In terms of geometry, this translates to a factor of 4 uncertainty in the estimated length along the line of sight, compared to the assumption that we are dealing with a spherical shell. This is, in principle, possible, but it is unlikely to exceed this factor of 4.

\subsection{Decisive tests}
In this section, we briefly discuss predictions of the model that can be verified with future observations. 

The most direct probe of the non-stationarity of the shell could come from high-resolution X-ray spectroscopy, by measuring the line velocities and line broadening at different positions across the shell. Double-horn lines separated by $\sim 2\,{\rm eV}$ (see Fig.~\ref{f:lprof}) would be a major indicator of the expanding shell.
Fine, eV-level X-ray spectroscopy in combination with a large grating is needed for this task. Mission concepts similar to \textit{LEM} \citep[e.g.,][]{2022arXiv221109827K,2023arXiv231016038K} would be ideal for this task. High angular resolution is not a must, so a less than $\sim 10$ degrees collimator could be sufficient.

Another possibility would be the detection of the ionization parameter gradient across the shell. On the angular scales of $\sim 10$ degrees considered in this study, the ionization parameter is already $\sim 3\times 10^{11}\,{\rm cm^{-3}s}$. One could try going down to sub-degree scales to identify the emergence of Fe~XVII lines, but this would require an accurate definition of the forward shock position. Yet another important step would be improving the constraints on the upstream plasma temperature. As discussed in Sect.\ref{s:cgm}, the main uncertainty comes from the clean removal of the foreground emission at low energies to identify lines/ions that are characteristic of the initial ionization state. As is clear from Fig.~\ref{f:rprof}, there are large-scale gradients in the distribution of the soft X-ray emission. A more accurate definition of the shock front and long observations of the selected region might mitigate the problem. 

Another difficult but important test could come from the identification of absorption lines (against background AGNs) characteristic of $\sim 0.6\,{\rm keV}$ plasma and sharing the same velocity structure. Once again, missions like \textit{LEM} could do the emission and absorption studies simultaneously, provided the right combination of the grasp and energy resolution.

\subsection{SE Bubble vs NPS}
\label{s:nps}
The east-west and north-south asymmetry of the eROSITA Bubbles is a subject of ongoing debate \citep[see, e.g.,][and reference therein]{2023CRPhy..23S...1L,2024A&ARv..32....1S}. One (out of many) possibilities is that the extremely bright NPS structure should be considered separately from the rest of the large-scale diffuse X-ray emission. For example, in  \cite{2024A&A...691L..22C}, the X-ray brightness of NPS is attributed to the high metallicity of the gas, which is uplifted by buoyancy from the region of active star formation well above the disk plane and shaped by the differential gas rotation. In this scenario, the remaining diffuse X-ray emission from the "proper" eROSITA Bubbles is the shock-heated low-metallicity gas that might be more symmetric. 

The similarities and differences in the surface brightness and spectral shape in different quadrants of the sky are clearly seen in the 3-color versions of the eROSITA all-sky maps (see Fig.~20 \citealt{2021A&A...656A.132S} and Fig.~1 in \citealt{2021A&A...647A...1P}).
To further illustrate this scenario using the SE shell as a reference, we used the publicly available MAXI maps \citep{2020PASJ...72...17N} in the 0.7-1.0~keV band. This map, after masking the brightest sources and smoothing, is shown in Fig.~\ref{f:maxi}. Two pairs of circles show the expected positions of X-ray-bright shells if the diffuse emission is perfectly symmetric and similar to the SE shell. Interestingly, one can see some X-ray emission inside these shells in all four quadrants. However, the correspondence is not perfect, and the mean surface brightness varies from one quadrant to another by a factor of at least 2. This is further illustrated in Fig.~\ref{f:s_vs_n}, where the radial profiles based on eROSITA data in the SE and NE shells are compared.

We conclude that treating the NPS separately from eROSITA bubbles remains a viable option, although extra observations are needed to support it further. For example, measuring line velocities, briefly discussed above, might provide extra leverage by demonstrating that the NPS emission is kinematically decoupled from the rest of the eROSITA bubbles emission. Another possibility is to measure abundances and individual element abundance ratios and show that several "varieties" of hot phases are present in the NPS region.

\subsection{Cosmic Rays and "PeV bump"}

\begin{figure}
\centering
\includegraphics[angle=0,trim=1cm 5.5cm 1cm 2.5cm,width=0.9\columnwidth]{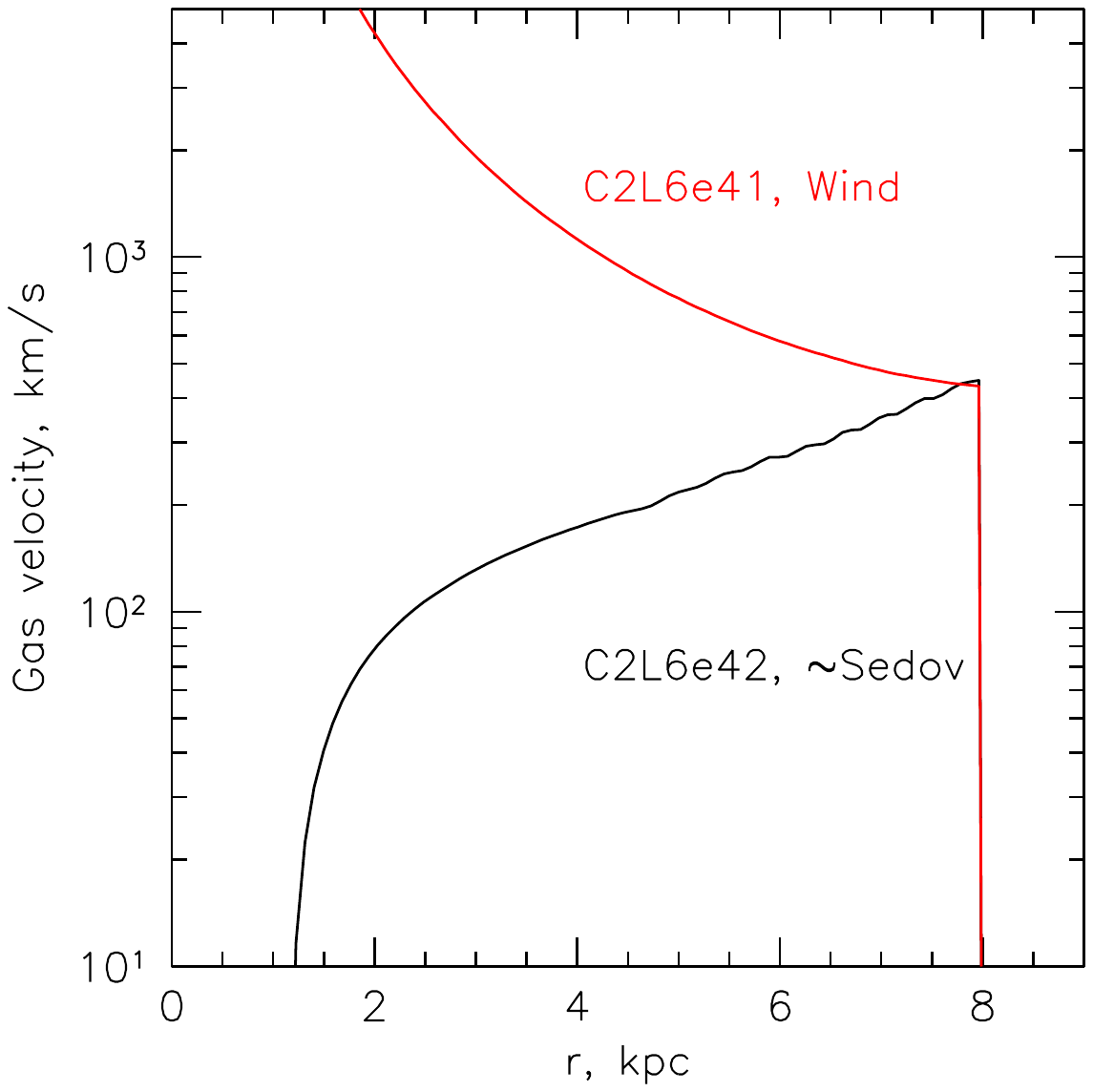}
\caption{Radial gas velocity profiles in two models. The black curve shows the case for a short, powerful explosion that evolves with time towards the Sedov-Taylor case. The red curve shows a steady-wind case. While the shock velocity is the same for both cases, the downstream behavior is different. The velocity decreasing with radius affects the distribution of cosmic rays downstream of the shock by advecting them closer to the shock.}
\label{f:velocity}
\end{figure}

The power of the transient outflow discussed above is of the order of 10$^{42} \ergs$ over several Myrs, exceeding the power of other extended Galactic sources like supernova remnants or accretion-powered microquasars, which are well-established particle accelerators. The residence time of the galactic GeV-energy cosmic rays (CRs) derived from the measured ratio of the Be/B ratio  ($^{10}$Be half-life is about 1.39 Myr) using a simplified leaky-box model is $\sim$ 15 Myr, and it can be longer in other models \citep[see e.g.][]{lipari2014lifetimecosmicraysmilky}. At higher CR energies, it decreases depending on particle energy, CR propagation model, and the halo size \citep[see e.g.,][]{2007ARNPS..57..285S,2019PhRvD..99f3012M}.  Therefore, the power and duration arguments suggest that the outflow that gave rise to the eROSITA bubbles might also act as an extended (a few kpc in size) accelerator affecting the CR population in the Galaxy.    While the detailed modeling of CR acceleration and propagation is beyond the scope of this paper, we briefly discuss a possible contribution of this source to the PeV CR fluxes at the "knee" of the galactic CR spectrum.

Recently, the LHAASO observatory measured 
the local (Solar System) cosmic ray proton spectrum in the energy range from 0.15 to 12 PeV. These measurements revealed a new proton-rich cosmic ray component \citep{2025SciBu..70.4173C} dubbed the "PeV bump" \citep{2026ApJ..1006..211A}. The LHAASO observatory also reported the hardening of the helium spectrum at about  1.1 PeV, followed by the spectrum softening at about  7 PeV \citep[][]{Cao_2026}. Helium nuclei fluxes become the dominant cosmic ray component at about 5 PeV.

We argue that these features may be understood in the scenario of diffusive shock acceleration to PeV-range energies in the low-metallicity plasma upstream of a large-scale shock.

First of all, the giant transient outflow that shaped the X-ray bubbles might efficiently accelerate Galactic cosmic ray ions up to PeV energies. Indeed, an MHD outflow with kinetic power $L_{\rm k}$ and velocity $V_{\rm f}$ can accelerate ions with charge $Z$ up to the energy $\sim 100\, Z \,\left (\zeta L_{\rm k}/10^{42} \ergs \right )^{0.5} \left (V_{\rm f}/1,000 \kms \right )^{0.5}$ PeV,  where $\zeta$ is the ratio of the magnetic luminosity to the kinetic power of the flow \citep[see e.g.,][]{1976Natur.262..649L,2009JCAP...11..009L}. The Hillas ion energy limit $E\sim  2 Z\,(V_{\rm f}/1,000 \kms)\, R_{\rm kpc}\, B_{\rm \mu G}$ PeV \citep{1984ARA&A..22..425H} by a flow of scale size $R_{\rm kpc}$, 
allows PeV proton acceleration in a 10-kpc-size outflow assuming the galactic halo toroidal magnetic field magnitude 
of $\gsim$ 0.3 $\mu$G \citep[][]{2024ApJ...966..240X,2025A&A...693A.284K}. This estimate was obtained from a general consideration of a proton interacting with an MHD flow carrying a frozen-in magnetic field. It provides an upper limit on the proton energy that can be achieved in ideal MHD-based acceleration mechanisms. Such energy corresponds to
the transition between the Galactic and extragalactic cosmic rays.

We suggest that the forward shock driven by the outflow can indeed accelerate nuclei to this limit via the diffusive shock acceleration mechanism.

The spectral index of a test particle momentum distribution is  $2(M^2 +1)/(M^2 -1) = 2.5$ for $M=3$ \citep[e.g.,][]{BE87}, assuming that the Alfvén speed is much lower than the outflow velocity.
Correction for the finite Alfvén speed upstream of the shock may soften the spectral index somewhat. 
The estimated maximal energy and the index of the power-law distribution make this outflow and its extended forward shock of potential interest as a galactic source of cosmic rays. If the magnetization parameter $\zeta \geq 0.01$, the maximal energies of a proton and helium nuclei accelerated by the outflow can reach the 10 PeV range. This energy, estimated from the general consideration of the MHD flow, is consistent with the estimate that follows from the nonlinear kinetic Monte Carlo modeling of the diffusive shock acceleration where the maximal energy scales with the ambient density $n$, shock velocity $\varv_{\rm sh}$ and the system size (free escape boundary) $S_{\rm esc}$ as $p_{\rm max} \propto n^{0.25}~\varv_{\rm sh}~S_{\rm esc}$ \citep{3inst2014}.

In the steady wind scenario C2L6e41 (the red curve in Fig. \ref{f:velocity}), the radial velocity of the outflow is decreasing with radius in the region bounded by the forward shock. In this case, the negative velocity gradient will keep shock-accelerated CRs closer to the
forward shock rather than letting them fill the entire downstream volume. Similarly, the distribution of particles escaping upstream is also confined to a region set by the diffusion coefficient and the shock velocity.

Given the large size (curvature radius) of the shock and an estimation of diffusion lengths, only CRs with energies $\gtrsim$PeV can escape and form a proton-rich component of the Galactic cosmic rays measured at Earth, while lower-energy particles will remain confined in the vicinity of the shock. Indeed, 
the accelerated particles will be localized in the vicinity of the forward shock in a region with size $l(E) \sim D(E)/\Delta u$, where $\Delta u \sim 1000 \kms$ is the characteristic velocity difference at the shock and across the downstream region, and $D(E)$ is the energy-dependent cosmic ray diffusion coefficient. To escape from the $\sim 10\,{\rm kpc}$ system, the diffusion coefficient has to be large,  $D(E) \gsim 10^{30} \diff$, which is plausible for PeV regime protons \citep[e.g.,][]{2007ARNPS..57..285S,2012SSRv..166...97A}. 

Therefore, the PeV regime particles accelerated by the large-scale shock 
in the 
CGM may constitute the galactic cosmic ray component,
designated Population~2 by \citet{2013FrPhy...8..748G}.  GeV-TeV particles will instead be confined in a few kpc vicinity of the forward shock. Electrons can be accelerated at the forward shock as well as protons, but their maximal energies are below TeV energies, being limited by the synchrotron-Compton losses. The inverse Compton emission of the accelerated leptons will form an extended gamma-ray source of low surface brightness well below that of the Fermi bubbles. This is because of the low electron injection rate into DSA acceleration from tenuous plasma in the shock upstream.

Finally, we note that the low metallicity of the gas upstream of the (eROSITA bubbles') shock would naturally lead to a proton-rich composition of the shock-accelerated cosmic rays. In particular, these cosmic rays may contribute to the "PeV bump", reported by LHAASO.

\subsection{Basic parameters and limitations}
Here, we reiterate that in Section 5.1, the characteristic values of the size and the age of FBs were derived by making several simplifying assumptions: 
\begin{itemize}
\item The geometry of the visible part of the southern FB is approximated with a 3D spherical shell (see Fig.5 and Fig.9). Other geometries are possible and, in fact, expected in numerical simulations \citep[e.g.,][]{2024A&ARv..32....1S}.
\item The net spectrum of the SEB was calculated by subtracting the background from adjacent regions. The advantage of this approach is that one can avoid fitting a multi-component model (source plus multiple background and foreground components) to the observed spectrum in the source region. We tried to make this procedure as robust as possible, but one can imagine a situation when an unrelated diffuse structure overlaps with the source (or background), biasing the results.
\item A spherically symmetric model in a uniform medium is used as a baseline model.
\item The shock radius is defined from the estimated distance to the shock boundary based on the analysis of the spectra.
\item The characteristic system age is taken directly from the simplified hydro model. One could instead use a value of $R/\varv$, which is related to the age of the outburst as $t_{\rm age}=\eta R/\varv$, where $\eta$ depends on the details of the time-dependent source of mass, momentum, and energy, and on the spatial distribution of the gas in the Galaxy.
An estimate of the latter effect can be made by considering the Sedov-Taylor problem for a power-law dependence of density with radius, i.e., $\rho(r)\propto r^{-m}$ \citep[e.g.,][]{1959sdmm.book.....S}. In this case, $t_{\rm age}=C\frac{R}{v}$, with $C=\frac{2}{5-m}=\frac{2}{5}, \frac{2}{4}, \frac{2}{3}$ for $m=0,1,2$, respectively. Therefore, for the $m=2$ profile, it takes a factor of $\sim 1.7$ times longer to reach the same radius with a given gas density (and the shock velocity), compared to the $m=0$ case. On the other hand, the largest contribution to the age is associated with the propagation over the last half of the current shock radius and, therefore, the correction factors are not very large. For the density profile $\rho(r)\propto r^{-1.5}$ \citep[see, e.g.,][]{2015ApJ...800...14M}, the correction factor will be $\sim 1.4$, and the values of $t_{\rm age}$ in Tab.~\ref{t:models} will be underestimated by a similar factor.

\item The metallicity of the gas was estimated from spectral analysis. However, different combinations of the ionization parameter, metallicity, and overall plasma temperature can produce largely similar X-ray spectra, especially when a limited energy range is considered. What drives the abundance low relative to the Solar values is the requirement to reproduce the observed flux and the ionization parameter. The gas density contributes to both quantities, while the abundance affects the flux. In terms of the goodness of the fit, $A\lesssim 0.1$ is preferred, but $A\sim 0.2$ reproduces the observed flux level slightly better for the adopted geometry. The derived value of $A$ depends directly on the adopted reference abundance table. Throughout the paper, the quoted abundances are relative to \cite{1989GeCoA..53..197A}. When tables of \cite{2003ApJ...591.1220L} or \cite{2009ARA&A..47..481A} are used, the resulting value of $A$ increases by a factor of $\sim 2$.
\end{itemize}

While any of these assumptions can undoubtedly affect the results, we believe that our results should be treated as an illustration of the broad consistency of the shock-driven NEI model with reported parameters and observations.

Alternatively, any detailed model (numerical simulations) can be confronted with the most basic and robust observables. Namely, 
the apparent angular dimension of the Southern EB (curvature radius $R\sim 32\,\rm deg$), effective downstream temperature ($kT\gtrsim 0.6\,\rm keV$) derived from the observed spectra\footnote{This value of temperature is revealed by the presence of Fe L-shell, Ne, and Mg lines.}, and the observed shell X-ray surface brightness in the 0.7-2~keV band ($I_{\rm X}\sim 3.2\times 10^{-16}\,{\rm erg\,s^{-1}\,cm^{-2}\,arcmin^{-2}}$).

\subsection{Western side of the eROSITA bubbles}
When this paper was under revision, a study focused on an extensive spectral analysis of the Western side of the eROSITA bubbles appeared on arXiv \citep{2026arXiv260502998Y}. The general spectral properties of the bubbles' X-ray spectra, namely the presence of spectral features characteristic of multi-temperature plasma, are broadly similar to those reported here. In the study of \cite{2026arXiv260502998Y}, the emphasis is placed on the widespread presence of the $kT\sim 0.2$ and $0.6\,{\rm keV}$ plasma components, while here we consider both components as signatures of the NEI plasma. The reported gas metallicity $Z=(0.2\pm 0.1)Z_{\odot}$ is consistent with that found in this study.
 
\section{Conclusions}

We analyzed the SE part of the eROSITA bubbles using the data from the SRG/eROSITA all-sky survey. Our conclusions can be summarized as follows:

\begin{itemize}
    \item The morphology of the SE portion of the eROSITA bubbles appears simpler compared to the NE part, where much brighter and less regular structures are observed. The shell-like morphology of the SE portion is consistent with the forward shock scenario. This interpretation and the required energy ($\sim 10^{56}\,{\rm erg}$) are consistent with the original interpretation of the eROSITA data in \cite{2020Natur.588..227P} as well as a broad class of models associated with the Galactic Center transient energy release \citep[e.g.,][]{1977A&A....60..327S}, see \cite{2024A&ARv..32....1S} for a recent review.
    \item In terms of the upstream gas temperatures, the values below 0.15 and above 0.25~keV are disfavored by the observed spectra (Fig.~\ref{f:kt0}), with the best-fitting values close to 0.2~keV. The latter value is affected by systematic uncertainties associated with the removal of foreground emission and variable photoelectric absorption for the region of such a large angular size. However, the bounds on temperature appear more robust. 
    \item The spectrum of the SE shell is consistent with the shock scenario, too (Sect.~\ref{s:spec}). This scenario requires a large ($\sim 10\,{\rm kpc}$) distance to the shell and a low abundance of the shock-heated gas ($Z/Z_\odot\lesssim 0.1$). The "local", e.g., at a distance of $\sim 100\,{\rm pc}$, scenario is excluded if the shell is indeed a forward shock. The key reason for this conclusion is the large ionization parameter $\tau$ derived from the spectra. 
    \item In the context of the shock model, the gas density at a distance of $\sim 10-12\,{\rm kpc}$ above the Galactic Disk is ~$n_e\sim 3\times 10^{-4}\,{\rm cm^{-3}}$ and the abundance of heavy elements in this gas is $Z\lesssim 0.1 \times Z_\odot$ (relative to the \cite{1989GeCoA..53..197A} abundances). Unlike constraints derived from the line-of-sight-integrated quantities, these are effectively in situ CGM measurements (Sect.\ref{s:cgm}). It is plausible that the abundance is higher closer to the Galactic disk, and these regions provide the dominant contribution to metal absorption lines and to the diffuse X-ray emission. Here we note that while in terms of the goodness of the fit, the model with $Z\sim 0.1 Z_\odot$ is preferred, a model with $Z\sim 0.2 Z_\odot$ reproduces the observed flux level slightly better without adjusting the normalization.  
    \item The best-fitting model predicts the current shock velocity $\sim 700\,{\rm km\,s^{-1}}$ and the age of the SE Bubble $\sim 7.5\,{\rm Myr}$. The true age might be larger when the gas density gradient is accounted for. For instance, in the Sedov-Taylor problem with a power law radial profile $\rho\propto r^{-1.5}$, the age will be a factor of 1.4 larger.
    \item Both the steady wind and the short outburst models can probably be adjusted to provide a reasonable approximation to the X-ray data considered here. The statistics accumulated by eROSITA allow differentiating between these models (Sect.~\ref{s:spec}), but uncertainties in the model assumptions are greater than the statistical ones. 
    \item The decisive consistency test of the forward shock models for the SE bubble would be a detection of lines of He- and H-like ions of Ne, Mg, Si, or Fe~XVII in emission and absorption. The latter requires collecting signals from many background AGNs (or selecting the brightest one in a favorable location). A match of the velocity patterns in emission and absorption would provide a robust and independent test of the model. The estimated column densities of the most promising ions are given in Table~\ref{t:columns}. The expected emission-line shape follows the "supernova-type" model with a prominent 2-peak structure (separated by $\sim 2\,{\rm eV}$ for the lines of Fe~XVII) for a line-of-sight going through the bubble (see Fig.~\ref{f:lprof}).  Fine, eV-level X-ray spectroscopy in combination with a large grasp is needed for such measurements, while the angular resolution is not critical. 
    \item The North/South asymmetry remains an interesting issue. The SE bubble clearly lacks extremely bright X-ray and radio structures, which are seen in the North and are bounded by NPS (see Appendix~\ref{a:nps_vs_sbub}). However, the NE region does have a faint X-ray emission at approximately the same location as in the SE bubble. Therefore, the models suggesting that NPS is a feature that comes on top of more symmetric GC-driven bubbles remain a viable option. In particular, the low  CGM abundance suggested by this study implies that any hot and metal-rich gaseous lamp can be prominent in X-rays on top of the low-metallicity GCM emission (Sect.~\ref{s:nps}).
    \item The size and the velocity of the SE-bubble shock propagating in the metal-poor plasma in the considered model make this shock a potential site for acceleration of  PeV cosmic rays in the Galaxy. This is particularly interesting in connection with the recently observed by LHAASO a proton-rich cosmic-ray component impinging on the Earth's atmosphere.
    
\end{itemize}

\begin{figure*}
\sidecaption
\includegraphics[angle=0,trim=1cm 9cm 1cm 9cm,frame,clip,width=12cm]{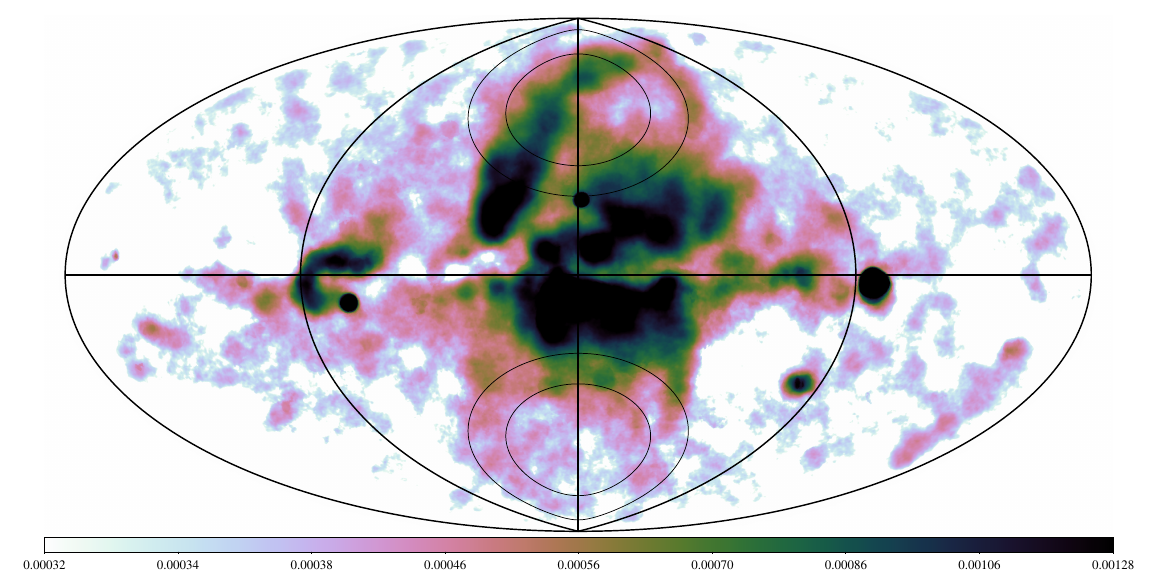}
\caption{MAXI 0.7-1 keV map with symmetric circular structures that mirror the SE Shell boundaries in the eROSITA maps (see Fig.~\ref{f:sbub_image}). The brightest compact sources were masked in the original MAXI map, and the residual image was smoothed with a 6 deg (FWHM) Gaussian.
}
\label{f:maxi}
\end{figure*}

\begin{acknowledgements}
We are grateful to our reviewer for their comments and suggestions.
This work is partly based on observations with the eROSITA telescope onboard the \textit{SRG} space observatory. The \textit{SRG} observatory was built by Roskosmos in the interests of the Russian Academy of Sciences represented by its Space Research Institute (IKI) in the framework of the Russian Federal Space Program, with the participation of the Deutsches Zentrum für Luft- und Raumfahrt (DLR). The eROSITA X-ray telescope was built by a consortium of German Institutes led by MPE and supported by DLR. The \textit{SRG} spacecraft was designed, built, launched, and is operated by the Lavochkin Association and its subcontractors. The science data are downlinked via the Deep Space Network Antennae in Bear Lakes, Ussurijsk, and Baikonur, funded by Roskosmos. 

The development and construction of the eROSITA X-ray instrument was led by MPE, with contributions from the Dr. Karl Remeis Observatory Bamberg $\&$ ECAP (FAU Erlangen-Nuernberg), the University of Hamburg Observatory, the Leibniz Institute for Astrophysics Potsdam (AIP), and the Institute for Astronomy and Astrophysics of the University of Tübingen, with the support of DLR and the Max Planck Society. The Argelander Institute for Astronomy of the University of Bonn and the Ludwig Maximilians Universität Munich also participated in the science preparation for eROSITA. The eROSITA data were processed using the eSASS/NRTA software system developed by the German eROSITA consortium and analyzed using proprietary data reduction software developed by the Russian eROSITA Consortium.

IK was supported by the Simons Foundation via the Simons Investigator Award to A. A. Schekochihin and by the COMPLEX project from the European Research Council (ERC) under the European Union’s Horizon 2020 research and innovation program grant agreement ERC-2019-AdG 882679. Cosmic ray acceleration modeling by AMB was supported by RSF grant No. 25-72-20007 (Multiscale nonlinear models of astrophysical sources of high energy radiation).

\end{acknowledgements}

\bibliographystyle{aa}
\bibliography{ref} 

\begin{appendix}

\section{Abundance and normalization degeneracy}
\label{a:abund}
\begin{figure}
\centering
\includegraphics[angle=0,trim=1cm 5.5cm 1cm 2.5cm,width=0.95\columnwidth]{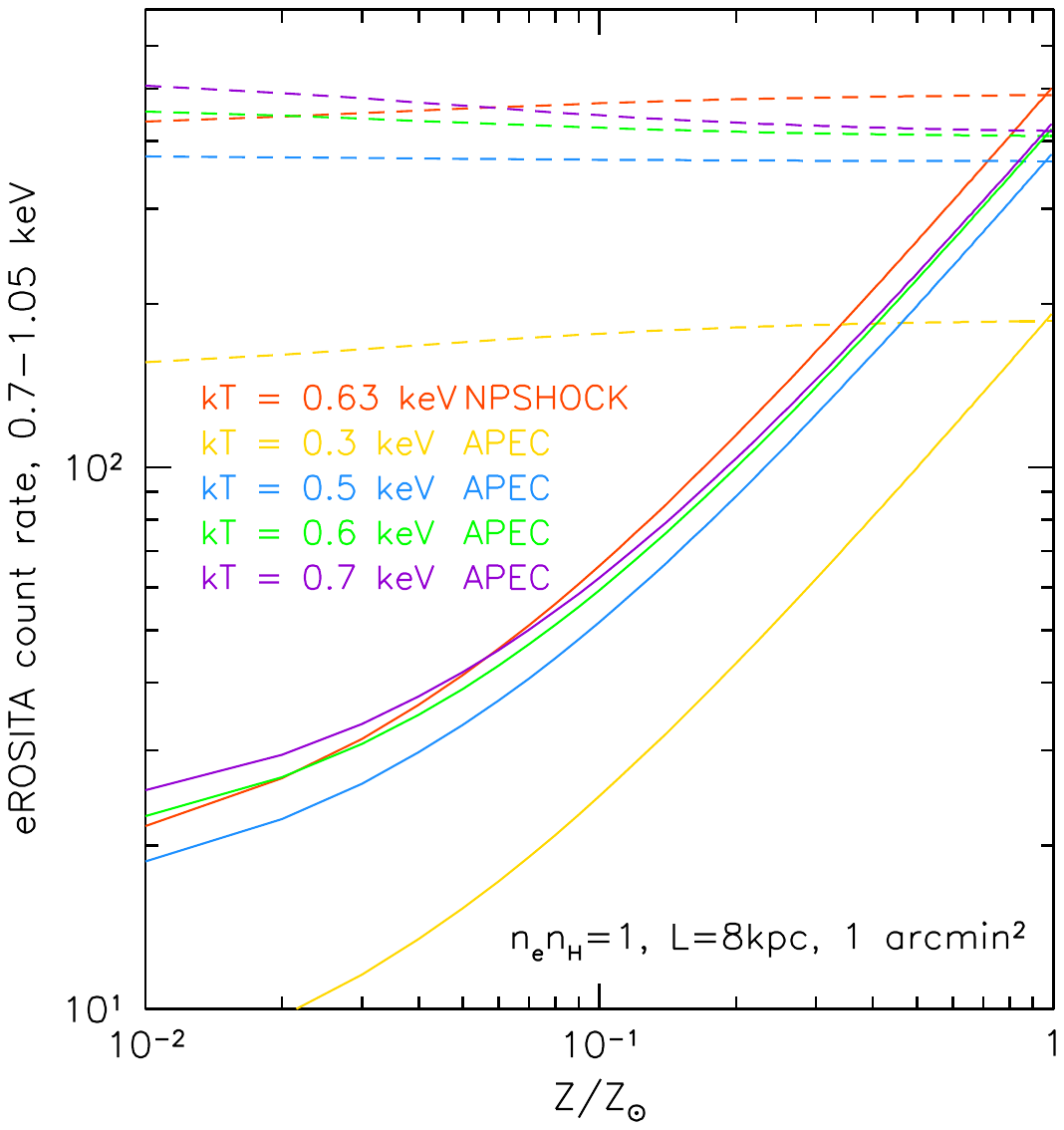}
\caption{eROSITA count rate in the 0.7-1.05~keV band (solid lines) as a function of metal abundance ($A=Z/Z_\odot$) for a set of models. All models have the same emission measure, corresponding to an 8~kpc-long slab with $n_en_p=1\,{\rm cm^{-6}}$ and the angular size of 1~sq.arcmin. The \texttt{NPSHOCK} model has the ionization parameter $\tau=2.8\times 10^{11}\,{\rm cm^{-3}s}$. 
The dashed lines show the same curves divided by $(A+0.04)$. For this set of models, this simple correction captures the metallicity dependence of the flux.}
\label{f:ix_ab}
\end{figure}

In Sec.\ref{s:spec}, we discuss the relation between the expected X-ray surface brightness and the gas emissivity in the 0.7-1.05~keV band (see eq.~\ref{e:ixobs}). While we solve this equation numerically, it is useful to see the dependence of the X-ray emissivity on the metal abundance explicitly. This is shown in Fig.~\ref{f:ix_ab} for a set of \texttt{APEC} models and the  \texttt{NPSHOCK} model (solid lines). It turns out that the dependence on the metallicity (in this band) is well captured by a simple function $\propto (A+0.04)$ over a relevant range of metallicities. 

\begin{figure}
\centering
\includegraphics[angle=0,trim=1cm 5.5cm 1cm 2.5cm,width=0.95\columnwidth]{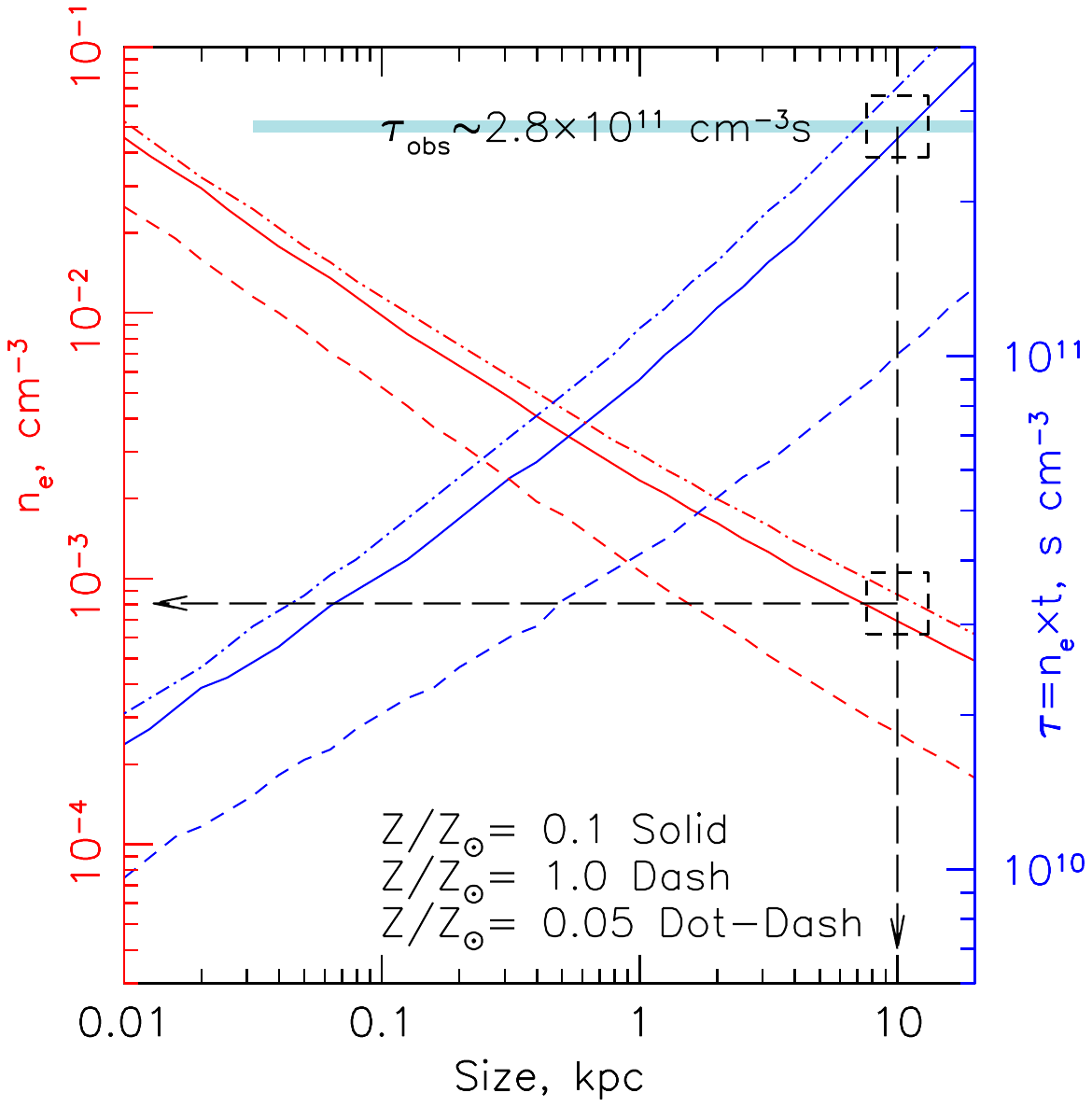}
\caption{Gas density $n_e$ downstream of the shock and the ionization time $\tau=n_et$ as a function of the size of the emitting region required to produce observed shell surface brightness $I_X$ in the 0.7-1.05~keV energy band. The dashed, solid, and dot-dashed lines correspond to the cases of metal abundance of 1, 0.1, and 0.05, respectively. 
The light-blue vertical bar shows the best-fitting value of the ionization parameter $\tau$. A viable solution should have an intersection of the observed and predicted values (blue lines) of $\tau$, shown by the top-left dashed box. Clearly, such solutions are possible only for low metallicity ($\lesssim 0.1 Z_\odot$) and large sizes of the emitting region $S\sim 10\,{\rm kpc}$ (see the vertical black arrow). Once the characteristic size is known, the downstream density can be determined, $n_e\sim (8-9)\times 10^{-3}\,{\rm cm^{-3}}$ (see the lower box and the horizontal arrow).  
While there are substantial uncertainties/assumptions in the procedure that led to this plot, they can not change the estimated size and density by a large factor.
}
\label{f:ix_size}
\end{figure}

The same \texttt{NPSHOCK} model is used to get initial estimates of the shell linear size, the gas density, and metal abundance, needed to get the observed surface brightness $I_X$ and ionization parameter $\tau$. This is illustrated in Fig.~\ref{f:ix_size}. From this figure, it follows that the linear size of the emitting region has to be $\sim 10\,{\rm kpc}$. Furthermore, an increase in the abundance requires a lower gas density (to get the same $\tau$) and, therefore, a larger linear size to get the same $I_X$.

\begin{figure*}
\centering
\includegraphics[angle=0,trim=1cm 5.5cm 1cm 2.5cm,width=0.66\columnwidth]{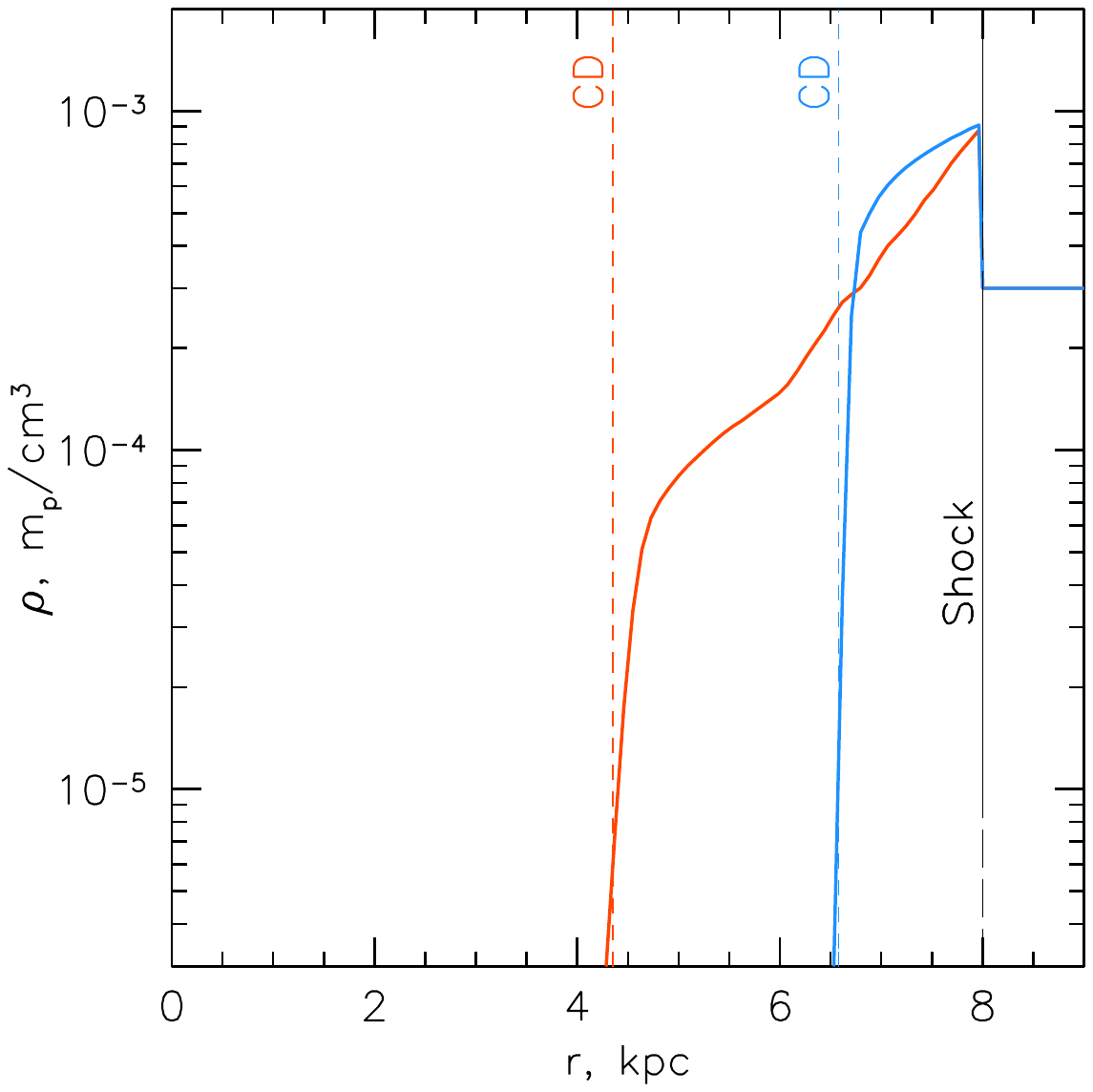}
\includegraphics[angle=0,trim=1cm 5.5cm 1cm 2.5cm,width=0.66\columnwidth]{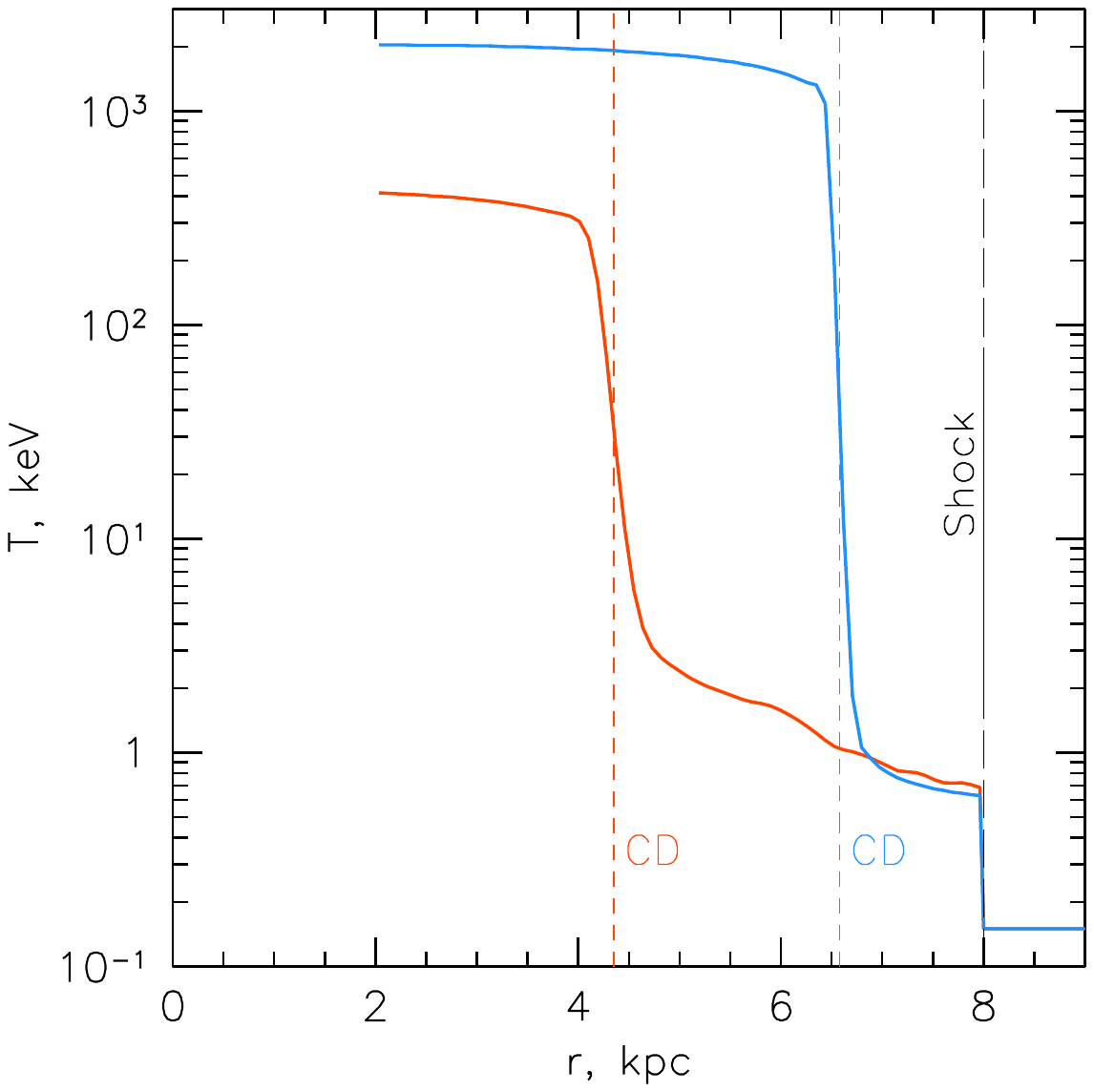}
\includegraphics[angle=0,trim=1cm 5.5cm 1cm 2.5cm,width=0.66\columnwidth]{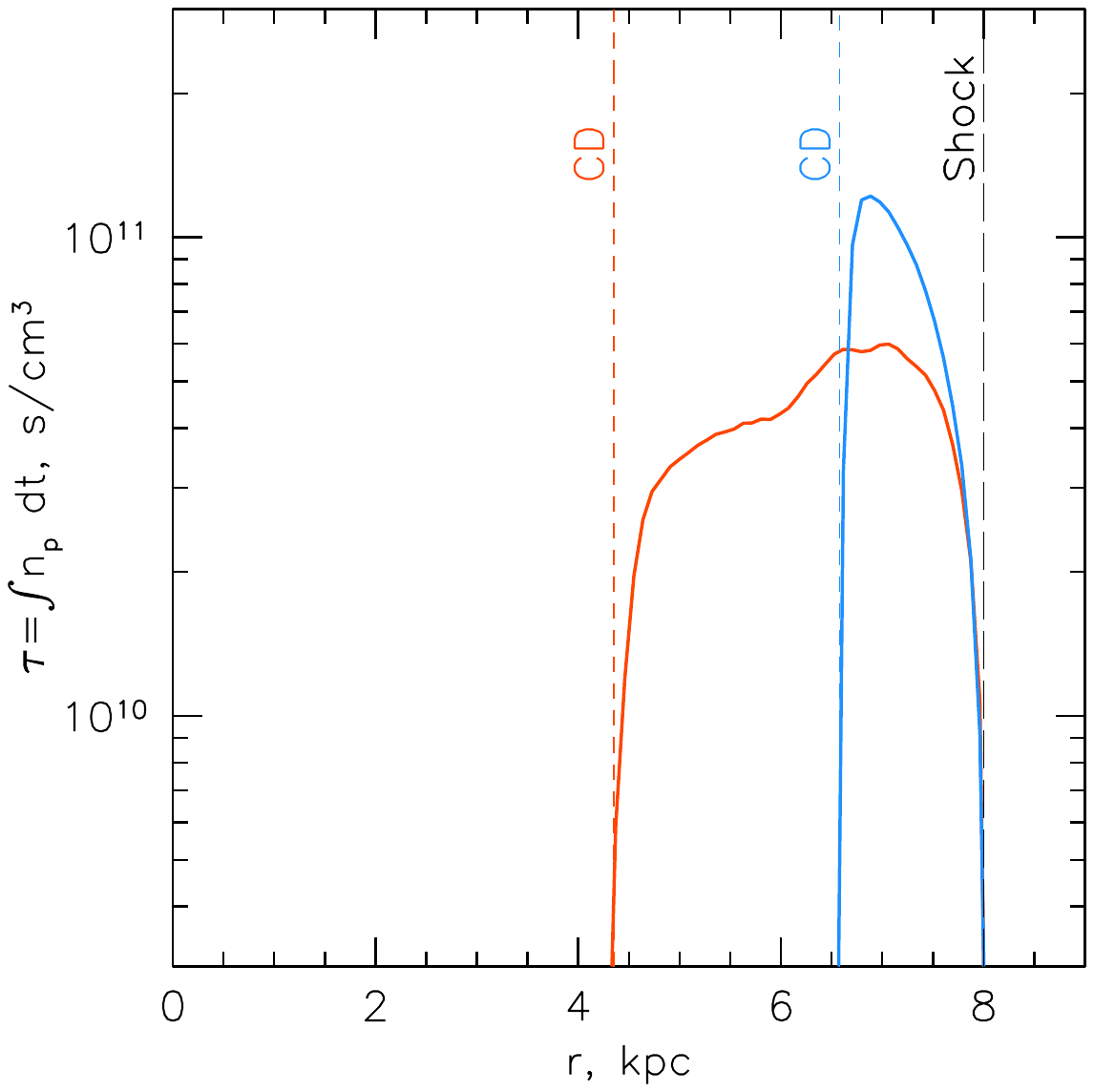}
\caption{Radial profiles of density, temperature, and ionization parameter in two spherically symmetric wind models, \texttt{C2L6e41} (blue) and \texttt{C2L6e42} (red). These models differ from each other only in the power of the central wind. In the more powerful, but shorter outburst scenario    (\texttt{C2L6e42}), the shock arrives earlier, leading to a smaller ionization parameter. Another difference between these two models is the position of the contact discontinuity. The maximal values of the ionization parameter are somewhat lower than in the best-fitting \texttt{NPSHOCK} model shown in Fig.~\ref{f:spec_npshock}. Nevertheless, the \texttt{C2L6e41} describes the observed spectrum reasonably well even without much fine-tuning.}
\label{f:hydro_model}
\end{figure*}

\section{Ionization and spectra}
\label{a:ib_spec}

\begin{figure}
\centering
\includegraphics[angle=0,trim=1cm 5.5cm 1cm 2.5cm,width=0.95\columnwidth]{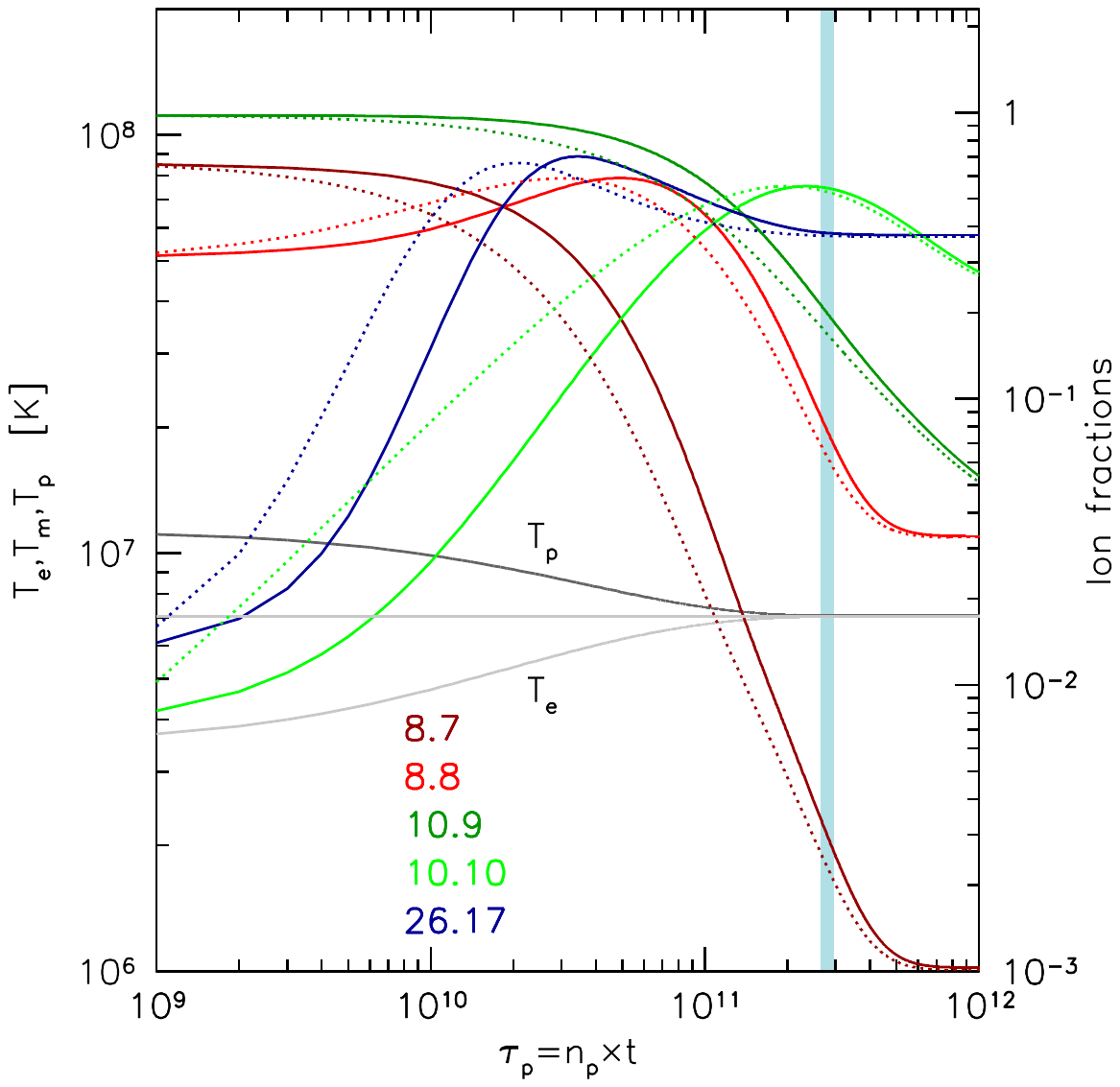}
\caption{Evolution of ionization fractions downstream of a plane shock with the Mach number $3.2$, propagating in the $T=0.15\,{\rm keV}$ gas with density $4\times 10^{-3}\,{\rm cm^{-3}}$. In this plot, we used $\tau_p=\int_0^t n_p dt$  rather than $\tau_e=\int_0^t n_e dt$ (for typical astrophysical plasma $\tau_p\approx 0.86\tau_e$).  The gray lines show the adopted evolution of electron and proton temperatures in the model when pure Coulomb collisions are responsible for the energy exchange between particles, while the horizontal line shows the case of $T_e=T_p$. The colored lines show the fractions for a subset of ions  (marked as "AtomicNumber.SpectroscopicSymbol"). Solid lines correspond to the $T_e=T_p$ case, while the dotted ones illustrate ion fractions when the electron temperature evolved with time due to Coulomb collisions. A similar model is used for calculations of ion fractions for 1D spherical models, taking into account the density and temperature evolutions in a given Lagrangian fluid element. }
\label{f:ib}
\end{figure}

Using the output of the simulations described above, it is straightforward to predict radial profiles of the proton and electron temperatures and the ionization state of each element/ion. When calculating the electron temperature profile, two extreme limits have been considered. In one limit, the electron and ion temperatures are equal and set to the expected mean plasma temperature, given the energy and mass density output of the simulations and setting the mean atomic weight $\mu=0.61$. In the other limit, the energy exchange between electrons and protons proceeds at the rate set by pure Coulomb collisions. The black and gray lines in Fig.~\ref{f:ib} illustrate the difference between these two cases. In general, for the ionization parameter larger than $10^{11}\, {\rm cm^{-3}s}$, electron and ion temperatures are equal, although in the outer layers the difference is not negligible.

To evolve the ionization balance, we use a set of rates for ionization and excitation from the CHIANTI database \citep{2007A&A...466..771D,2021ApJ...909...38D}. The typical evolution of the selected ion fractions for O~VII, O~VIII, Ne~IX, Ne~X, and Fe~XVII is shown in Fig.~\ref{f:ib} with the colored lines. Here, we used a plane-parallel shock with a Mach number of $3.2$ propagating through the $0.15$~keV gas for illustration.  The solid lines correspond to the case when $T_e=T_i$, while the dotted lines illustrate the case when the electron temperature evolves with time due to Coulomb collisions. The same processes are included in simulations of the spherical shock associated with the Southern eROSITA bubble.

Finally, to calculate the X-ray emissivity in each radial shell, we use the ionization fractions for each shell and the emissivities from the ATOMDB database  \citep[version 3.1.3,][]{2012ApJ...756..128F}. Once the emissivities are calculated, the projected spectra are calculated as a function of projected radius in units of the shock radius, i.e., $R_p/R_s$, which are compared with the observed spectra. 

\section{Column density of ions and the emission line broadening}
\label{a:lines}
\begin{figure}
\centering
\includegraphics[angle=0,trim=1cm 5.5cm 1cm 2.5cm,width=0.95\columnwidth]{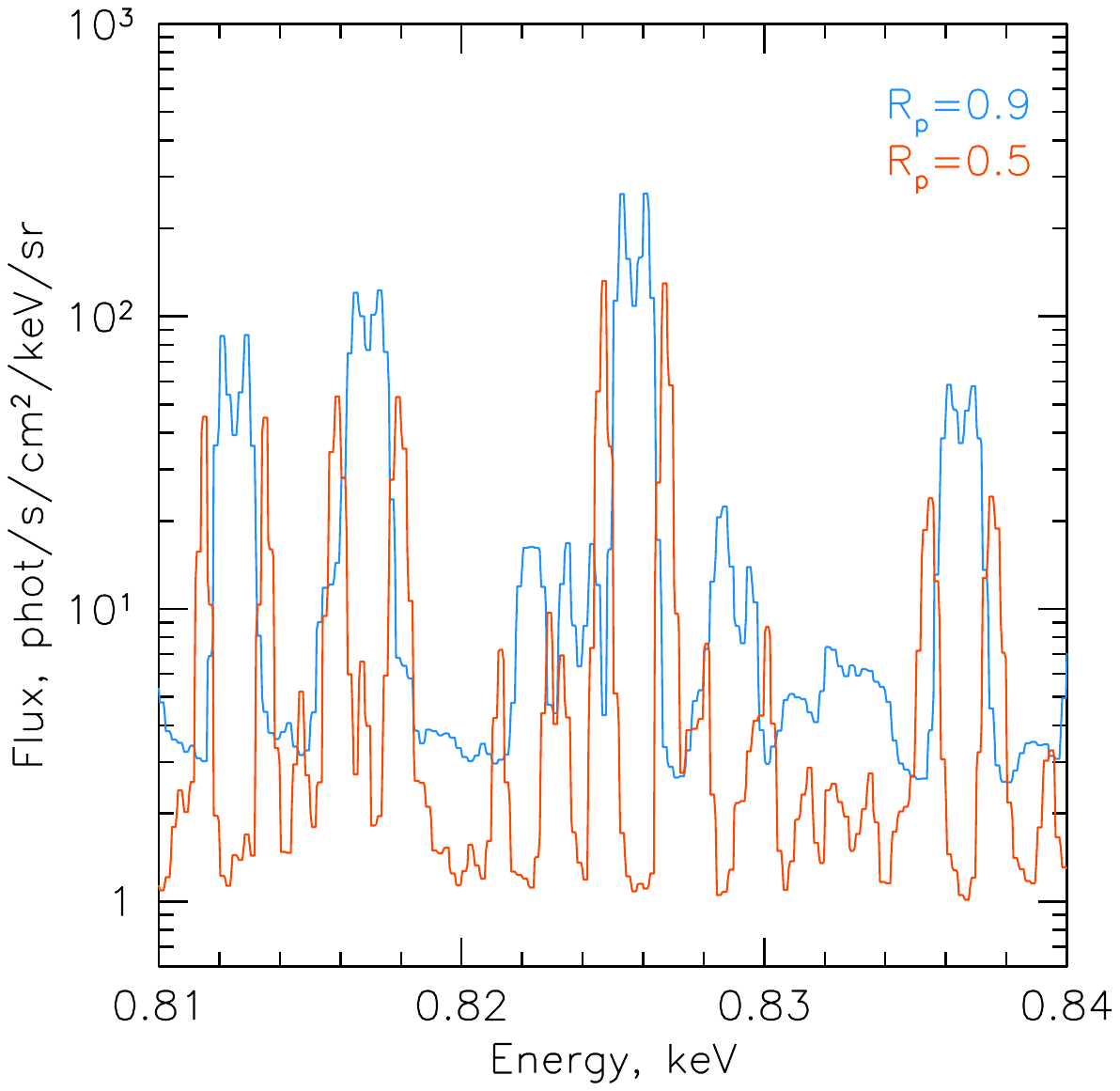}
\caption{Emission line profiles at two projected radii for the model \texttt{C0L6e41}. This model is similar to \texttt{C2L6e41}, except that the temperatures of electrons and ions are always equal. The projected radii are in units of the forward shock radius. The brightest line in the center corresponds to the Fe~XVII line at 826 eV. At $R_p=0.9 R_{s}$, the line of sight is largely edge-on, and the broadening is modest. On the contrary, at $R_p=0.5 R_{s}$, the line splits into two well-separated [$\sim 2\,{\rm eV}$] narrow components due to expansion of the shell.  }
\label{f:lprof}
\end{figure}

Measuring absorption column densities of ions characteristic of a particular CGM temperature, in combination with the detection of the same lines in emission, is a powerful tool for gas diagnostics.
In the case of a moving/expanding medium, both the number of ions and the line broadening due to gas motions might be important. In Table~\ref{t:columns} we list the column densities of the most important ions for the line of sight crossing the shell. The line shapes at two projected distances from the shell center are shown in Fig.~\ref{f:lprof}. A typical contribution of the gas motions (at $E\sim 0.8\,{\rm keV}$) to the lines width is $\sim 2\,{\rm eV}$.

The overall scheme of observing lines in emission and absorption is discussed in, e.g.,  \cite{2019MNRAS.482.4972K}, \cite{2023MNRAS.523.1209C}.  One needs to use telescopes with a large grasp to measure diffuse emission and observe one (or preferably many) strong sources in the background in the same (or similar) region with a high-energy-resolution instrument. The most promising would be using a telescope that combines these two properties, such as \textit{LEM} \citep{2022arXiv221109827K}.

\section{SE-wedge vs NE-shell}
\label{a:nps_vs_sbub}
\begin{figure}
\centering
\includegraphics[angle=0,trim=1cm 5.5cm 1cm 2.5cm,width=0.95\columnwidth]{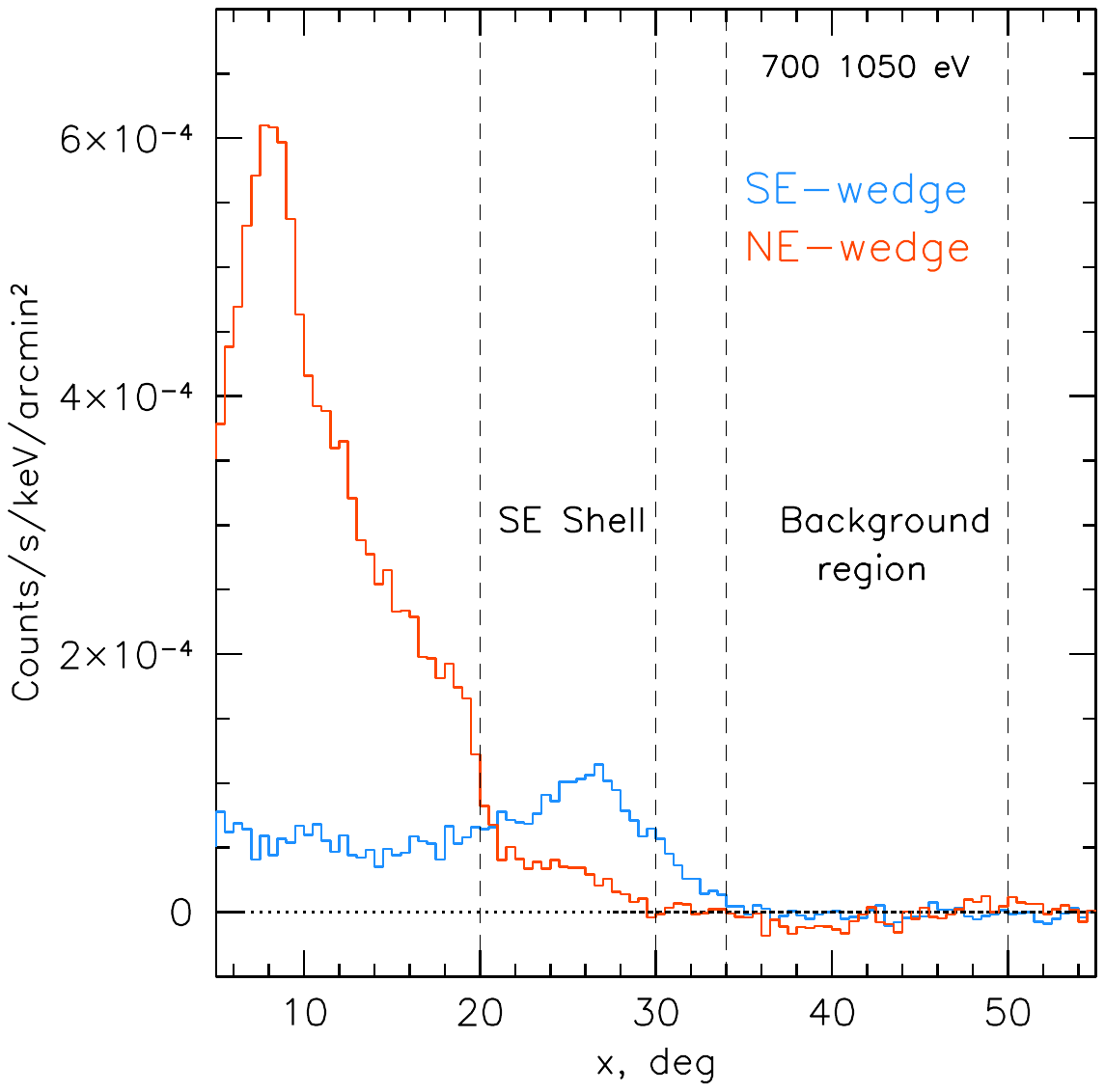}
\caption{Comparison of the radial X-ray surface brightness profiles in the SE Bubble and in the similar region in the NE region (see Fig.~\ref{f:sbub_image}). The sky background was subtracted from both profiles.  In the radial range corresponding to the SE shell (20-30 degrees from the reference point), the excess X-ray flux is seen in both SE and NE profiles, albeit it is a factor of 2-3 fainter in the North. At smaller radii, the NPS region on the North outshines the SE profiles by a factor of $\sim 6$. This leaves open a question on the correspondence (or the lack of) between the structures on the opposite sides of the Galactic Center.}
\label{f:s_vs_n}
\end{figure}

 In this section, we directly compare the radial profiles of the SE Bubble and the same region in the NE. Fig.~\ref{f:s_vs_n} shows these profiles in the 0.7-1.05~keV band. The astrophysical and detector backgrounds have been subtracted. For clarity, the vertical lines divide the radial range into three regions: "background" (34-50 degrees), "SE shell" (20-30 degrees), and the inner part (less than 20 degrees). The "NPS" emission dominates the latter region in the North, where the surface brightness is an order of magnitude higher than in the South. In the North, some excess emission is present in the "shell" region, albeit a factor of $\sim 2$ fainter.  
 
\begin{table}[]
\caption{Expected column densities of the shell for various ions }
    \centering
    \begin{tabular}{l|r}
    \hline
    \hline
    Ion & Column density $\rm cm^{-2}$ \\
    \hline
    \hline
    HII   &1.3E+19 \\
O~VII    &2.3E+14   \\
O~VIII   &4.7E+14  \\
Ne~IX   &1.0E+14  \\
Ne~X    &5.8E+13   \\
Mg~XI   &4.0E+13  \\
Mg~XII  &6.4E+12  \\
Si~XIII   &3.4E+13  \\
Si~XIV   &1.2E+12   \\
Fe~XVII &2.7E+13   \\
           \hline
       \hline
    \end{tabular}
    \tablefoot{Column densities are calculated for a subset of ions by integrating the number density of ions in the \texttt{C2L6e41} model, along the line of sight crossing the shell at projected distances from the center (in units of the shock radius) where this column density is maximal (typically $\sim 0.8-0.95$ of the shock radius). The HII column density is quoted to illustrate the total proton column density. The abundance of heavy elements in this model is 0.1 Solar.     }
    \label{t:columns}
\end{table}

\section{Uniformity of the sky 0.7-2 keV brightness at high latitudes}
\begin{figure}
\centering
\includegraphics[angle=0,trim=1cm 5.5cm 1cm 2.5cm,width=0.95\columnwidth]{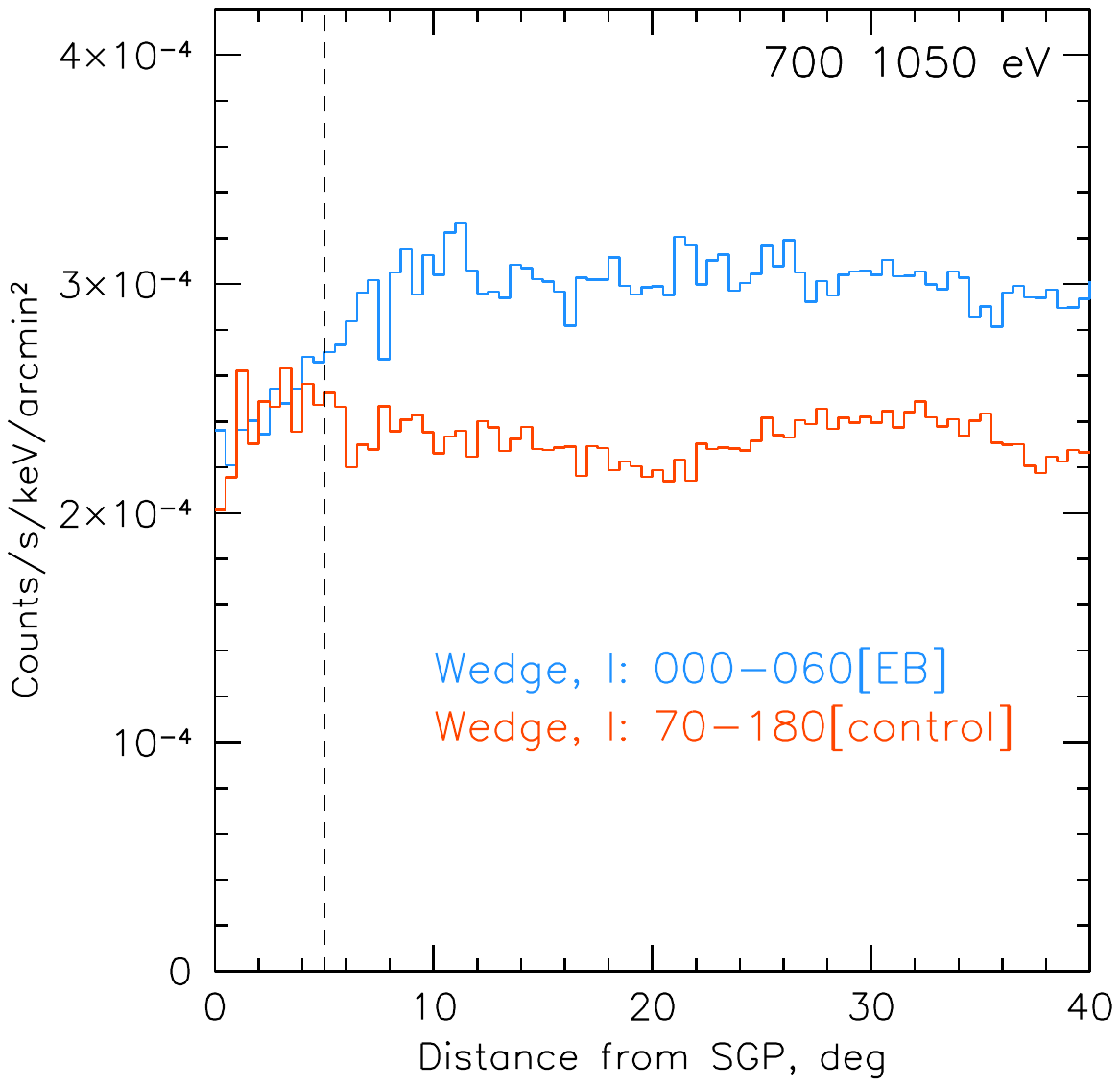}
\caption{Relative uniformity of the sky brightness in the 0.7-1.05~keV band at high latitudes. The sky brightness was calculated in two wedges with vertices at the South Galactic Pole (SGP). One wedge covers the range of galactic longitudes between $l=0$ and $l=60$ degrees (blue, encapsulating the SE-shell), while the other one spans from $l=70$ to $l=180$ (red). The sky brightness profiles are plotted as a function of distance from the SGP, corresponding to galactic latitudes from $b=-90^\circ$ to $b=-50^\circ$.  The vertical dashed line shows the top boundary of the SE-shell, some 5 degrees from SGP. In this band, the variations of the sky brightness are modest, and, therefore, regions outside the SE-shell could be used for the determination of the background contributing to the total emission from the SE-shell region. }
\label{f:07-2b}
\end{figure}

The possibility of removing and subtracting the sky background from the SE-shell is an important ingredient of the analysis. Here, we do additional tests to show that above 0.7~keV, i.e., above the OVIII~$\rm Ly\alpha$ line, the sky brightness is rather uniform at high galactic latitudes $b<-50^\circ$.
This is, in fact, expected, given that much of the Milky Way's diffuse emission is at lower energies. Also, the photoelectric absorption is less important at these energies, especially for small column densities characteristic of high latitudes. This is illustrated in Fig.~\ref{f:07-2b}. While some variations are present, their amplitudes are small enough to justify the evaluation of the SE-shell contribution to the total spectrum by subtracting the background spectrum measured in adjacent regions.

\section{SE-shell spectrum 0.7-2 keV}
In this section, we repeat fitting the spectra with a subset of models, but using only the 0.7-2 keV portion of the spectrum. The best-fitting parameters for the \texttt{APEC}, \texttt{NPSHOCK}, and our \texttt{C2L6e41} model with three different metal abundances (relative to the \cite{1989GeCoA..53..197A} abundances) are shown in Table~\ref{t:models07}.

First of all, the \texttt{APEC} model gives a higher temperature of $0.6\,{\rm keV}$ compared to $0.195\,{\rm keV}$ in Table~\ref{t:models}. This behavior is expected, since a one-temperature plasma in collisional ionization equilibrium can not describe the non-equilibrium spectrum.

Secondly, the \texttt{NPSHOCK} model yields almost the same plasma temperature and the ionization parameter as in Table~\ref{t:models}. However, the best-fitting abundance is now $\sim 0.19\pm0.4$, i.e., almost twice as high as in Table~\ref{t:models}.

In the case of the \texttt{C2L6e41} model, the lower abundance ($A\sim 0.1$) remains the preferred solution from the point of view of the $\chi^2$. However, we note that the most self-consistent solution, when the normalization predicted by the models coincides with the observed flux, is achieved for $A\sim 0.2$.  

Increasing the abundance to $A\sim 0.3$ further increases the value of  $\chi^2$ and overpredicts the observed flux.

\begin{table}[]
\caption{Models approximating the background-subtracted shell spectrum in the 0.7-2~keV band.}
    \centering
    \begin{tabular}{l|l|l}
    \hline
    \hline
    Model & Parameters & $\chi^2$, 168 bins \\
    \hline
    \hline
        \texttt{APEC} & $kT=0.604 \pm  0.0084 \,{\rm keV}$ & 195.5 \\
        & $A=0.16 \pm 0.03$ & \\
         \hline
        \texttt{NPSHOCK} & $kT_a= 0.69 \pm  6.4\times 10^{-2} \,{\rm keV}$ & 153.0 \\
       &  $kT_b= 0.15 \,{\rm keV; frozen}$ & \\
       & $\tau=2.2\times 10^{11}  \pm 5.6\times 10^{10}\,{\rm cm^{-3}s}$ & \\
        & $A=0.19 \pm 0.039$ & \\   
        \hline
        \texttt{C2L6e41}  & $R=\left [0.7R_s : R_s\right ]$; frozen & \\
        Steady wind & $A=0.1$; frozen & \\
        & Normalization; free ($1.78$)  & 157.8 \\
          \hline
         \texttt{C2L6e41}  & $R=\left [0.7R_s : R_s\right ]$; frozen & \\
        Steady wind & $A=0.2$; frozen & \\
        & Normalization; free ($0.98$)  & 181.8 \\
          \hline
        \texttt{C2L6e41}  & $R=\left [0.7R_s : R_s\right ]$; frozen & \\
        Steady wind & $A=0.3$; frozen & \\
        & Normalization; free ($0.68$)  & 198.7 \\
          \hline
       \hline
    \end{tabular}
    \tablefoot{Same as Tab.~\ref{t:models} but using only the data in the 0.7-2~keV band, where the background and low-energy photoelectric absorption are less important.  The metal abundances are relative to \citealt{1989GeCoA..53..197A}.
}
    \label{t:models07}

\end{table}

\section{Constraints on the upstream temperature}
\label{a:kt0}
\begin{figure}
\centering
\includegraphics[angle=0,trim=1cm 5.5cm 1cm 2.5cm,width=0.95\columnwidth]{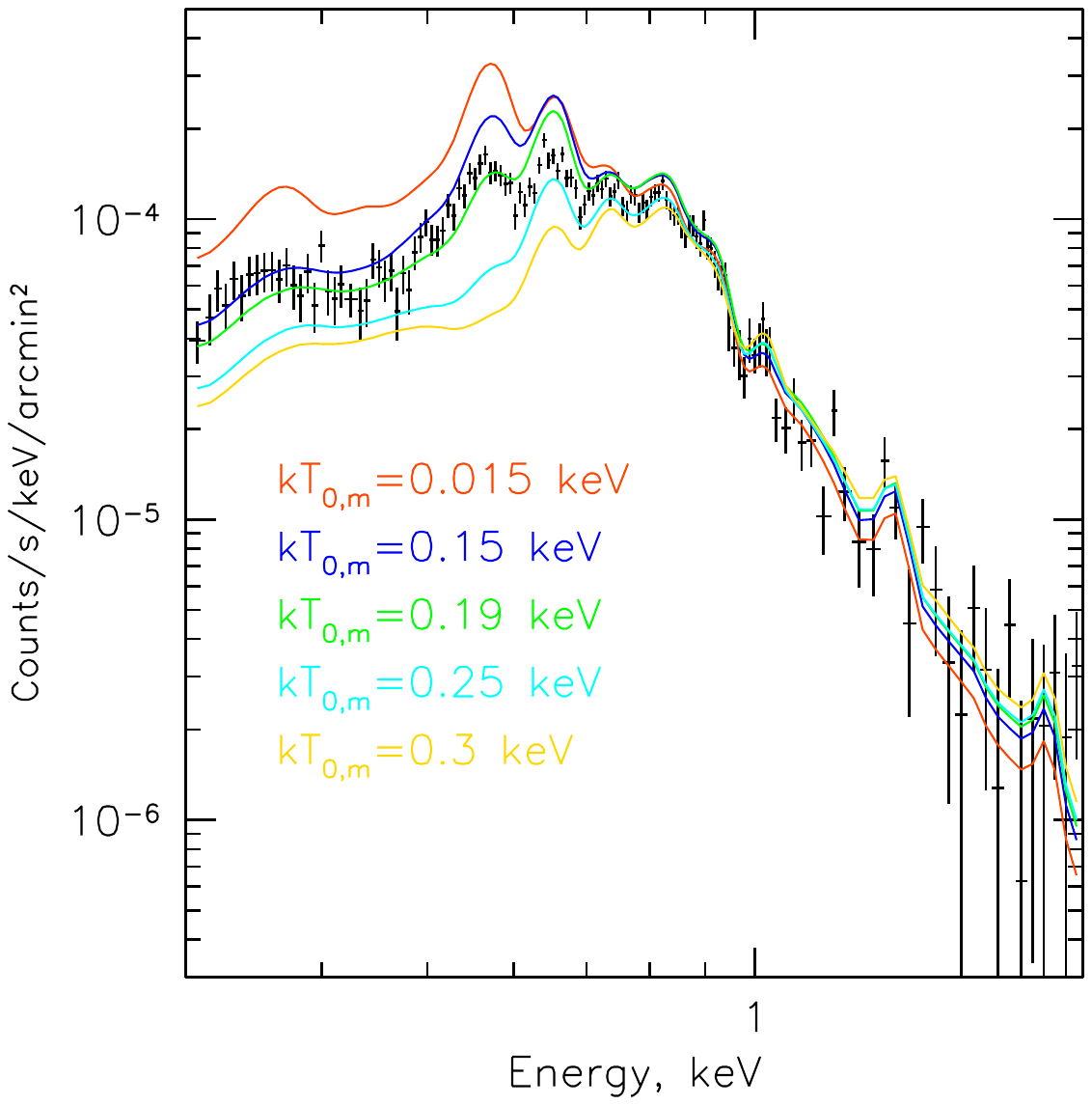}
\caption{Constraints on the upstream gas temperature. The black points show the same spectrum as in Fig.~\ref{f:spec_npshock}. The colored curves correspond to our baseline model \texttt{C2L6e41} for different upstream temperatures $T_{0,m}$. The red curve corresponds to the rather cold plasma (0.015~keV) and clearly predicts an OVII line that is significantly stronger than the OVIII line, which is not observed. On the contrary, very hot plasma (0.25 or 0.3~keV) lacks the OVII line completely, in contrast with the data. Two intermediate solutions with temperatures 0.15 and 0.19~keV look qualitatively acceptable, setting the range of the plausible upstream temperatures. }
\label{f:kt0}
\end{figure}

In this section, we illustrate the impact of the upstream temperature $kT_{0,m}$ on the shell spectrum. The upstream plasma is in CIE at the corresponding temperature. We run our baseline model \texttt{C2L641} with the same parameters except for the values of $kT_{0,m}$, which we set to $0.015,\, 0.15,\,0.19,\,0.25,$ and $0.3$~keV and calculate the shell spectra when the shock reaches the same distance from the source. The spectra predicted by this model are shown in Fig.~\ref{f:kt0}. For comparison, the observed (background-subtracted) spectrum is shown with black crosses. Clearly, the greatest changes are associated with 
OVII line, which is overproduced for low values of $kT_{0,m}\lesssim 0.15\,\rm keV$ and is underproduced for the higher values of $kT_{0,m} \gtrsim 0.20\,\rm keV$. This can be considered as an efficient in situ thermometer of the halo gas temperature. However, the accuracy of the background subtraction might be lower at these relatively low energies. For these reasons, we consider the interval between 0.15 and 0.2~keV as a plausible range for the upstream temperature and use $kT_{0,m}=0.15\,\rm keV$ in the model. In terms of the $\chi^2$, the model with  $kT_{0,m}=0.19\,\rm keV$ is preferred (from the considered set), followed by the model with $kT_{0,m}=0.15\,\rm keV$. All other models have much larger $\chi^2$, as is also evident from the spectra shown in Fig.~\ref{f:kt0}. We further illustrate this statement in Tab.~\ref{t:t0}, which shows the values of the $\chi^2$ for a subset of \texttt{C2L6e41} models with different upstream temperatures.

For comparison, Table~\ref{t:taab} shows the impact of the initial electron temperature for the \texttt{NPSHOCK} model and the abundance ratios from \citealt{2003ApJ...591.1220L}. In this model, this temperature (the so-called $kT_{\rm b}$ parameter) corresponds to the electron temperature downstream of the shock. As a result, the changes in the initial temperature do not strongly affect the shape of the spectrum. On the contrary, the higher abundance of oxygen and iron in the table of \citealt{1989GeCoA..53..197A} affects both the normalization and the derived metallicity.

\begin{table}[]
\caption{Sensitivity of the model spectra to the upstream gas temperature $kT_{0}$. 
}
    \centering
    \begin{tabular}{l|l}
    \hline
    \hline
    $kT_{0}$, keV & $\chi^2$, for 235 bins \\
    \hline
    \hline
    0.10 & 661.1 \\
    0.15 & 482.3 \\
    0.20 & 466.1 \\
    0.25 & 846.6 \\
           \hline
       \hline
    \end{tabular}
    \tablefoot{The data in the 0.3-2~keV band are used. The \texttt{C2L6e41} model was generated for a small grid of temperatures and compared with the observed spectrum. The metal abundances are fixed to $0.1$ relative to \citealt{1989GeCoA..53..197A}. The normalization of the model was allowed to vary. The table provides the value of the $\chi^2$ for each value of $kT_{0}$. As expected (see Fig.~\ref{f:kt0}), the models with $kT_{0}=0.15$ and 0.2 keV perform better than others.}
    \label{t:t0}
\end{table}

\begin{table}[]
\caption{\texttt{NPSHOCK} model fit parameters for a small grid of initial downstream electron temperatures using the Solar abundance ratios from \cite{2003ApJ...591.1220L}. 
}
    \centering
    \begin{tabular}{l|l|l|l|l}
    \hline
        \hline

    $kT_{\rm b}\,\rm, keV$ & $kT_{\rm a}\,\rm, keV$ & $A$ & $N$ & $\chi^2$ (231 d.o.f) \\
    \hline          
        0.10 & 0.66 & 0.19  & $6.4\times 10^{-7}$ &255\\
        0.15 & 0.66 & 0.19  & $6.4\times 10^{-7}$ &255\\
        0.15 & 0.64 & 0.10  & $7.3\times 10^{-7}$ &269 AG89 \\
        0.20 & 0.66 & 0.19  & $6.4\times 10^{-7}$ &256\\
        0.25 & 0.67 & 0.19  & $6.4\times 10^{-7}$ &258\\
          \hline
       \hline
    \end{tabular}
    \tablefoot{ This exercise aims to illustrate the level of uncertainties associated with the "latent" parameters (such as the initial electron temperature $kT_{\rm b}$). Clearly, the \texttt{NPSHOCK} model is not sensitive to the value of $kT_{\rm b}$. We reiterate here that $kT_{\rm b}$ in the \texttt{NPSHOCK} model corresponds to the electron temperature immediately downstream of the shock, while in our wind models, the initial temperature $kT_0$ characterizes the ionization state of the gas upstream of the shock. 
    The model parameters obtained for the Solar abundance ratios from \cite{2003ApJ...591.1220L} can be compared with the \texttt{NPSHOCK} entries in Tab.~\ref{t:models} and ~\ref{t:models07}, where abundance ratios from \cite{1989GeCoA..53..197A} are used. For convenience, we show corresponding parameters for $kT_{\rm b}=0.15\,\rm keV$ in the line labeled with "AG89". For brevity, statistical uncertainties are not quoted. 
}
    \label{t:taab}
\end{table}

\end{appendix}

\label{lastpage}
\end{document}